\documentclass[11pt,a4paper]{article}
\usepackage{jheppub}
\usepackage[normalem]{ulem}
\usepackage{graphicx}
\usepackage{epstopdf}
\usepackage{multirow}
\usepackage{rotating}

\def\vec#1{\mathbf{#1}}
\def\fKpi{f_+^{K\pi}}

\newcommand{\fpzero}{\fKpi(0)}

\newcommand{\be}{\begin{equation}}
\newcommand{\ee}{\end{equation}}
\newcommand{\bi}{\begin{itemize}}
\newcommand{\ei}{\end{itemize}}
\newcommand{\fplus}{f^{K\pi}_+}

\newcommand{\fminus}{f^{K\pi}_-}
\def\ltorder{\mathrel{\raise.3ex\hbox{$<$}\mkern-14mu
 \lower0.6ex\hbox{$\sim$}}}
\def\gsim{\mathrel{\rlap{\lower4pt\hbox{\hskip1pt$\sim$}}
    \raise1pt\hbox{$>$}}}                

\begin{document}

\title{The kaon semileptonic form factor in $N_f=2+1$ domain
	wall lattice QCD with physical light quark masses}

\author{P.A.~Boyle$^a$,
{N.H.~Christ$^b$},
{J.~M.~Flynn$^c$},
N.~Garron$^d$,
C.~Jung$^e$,
{A.~J\"uttner$^c$},
{R.~D.~Mawhinney$^b$},
{D.~Murphy$^b$},
{C.~T.~Sachrajda$^c$},
{F.~Sanfilippo$^c$},
{H.~Yin$^{b,\dagger}$}\\[2mm]
RBC and UKQCD Collaborations\let\thefootnote\relax\footnotetext{$^\dagger$Current address: 1600 Amphitheatre Parkway, Mountain View, CA 94043}\\[-2mm]}

\affiliation{$^a$School of Physics \& Astronomy, University of Edinburgh, EH9 3JZ, UK}
\affiliation{$^b$Physics Department, Columbia University, New York, NY 10027, USA}
\affiliation{$^c$School of Physics and Astronomy, University of Southampton,  Southampton, SO17 1BJ, UK}
\affiliation{$^d$School of Computing and Mathematics and Centre for Mathematical Science, 
Plymouth University, Plymouth PL4 8AA, United Kingdom}

\affiliation{$^e$Brookhaven National Laboratory, Upton, NY 11973, USA }

\abstract{
 We present the first calculation of the kaon semileptonic 
 form factor 
 with sea and valence quark masses tuned to their physical values
 in the continuum limit of 2+1 flavour domain wall lattice QCD.
{ We analyse a comprehensive
  set of simulations at the phenomenologically convenient point of
  zero momentum transfer in large physical volumes and for two
  different values of the lattice spacing.}
 Our prediction for the form factor is
  $\fpzero=0.9685(34)(14)$ where the first error is 
 statistical and the second error systematic. This result can be combined
 with experimental measurements of $K\to\pi$ decays 
 for a determination of the CKM-matrix element for which we 
 predict $|V_{us}|=0.2233(5)(9)$ where the first error is from 
 experiment and the second error from the lattice computation. 
}


\maketitle
\section{Introduction}
In the Standard Model (SM) the unitary Cabibbo-Kobayashi-Maskawa (CKM)
matrix~\cite{Cabibbo:1963yz,Kobayashi:1973fv}
parameterises flavour
changing processes and  provides  the only source of CP-violation in 
the quark flavour sector. 
Determining its elements from a combination of theory predictions for
flavour-changing SM processes and the corresponding experimental measurements
allows for testing its unitarity and consistency.
Any deviation from the expected behaviour would signpost {\it new physics}.

{In this work we present a new lattice QCD
computation of the hadronic contribution to the semileptonic
$K\to\pi$ decay, the vector form factor $f_+^{K\pi}(0)$,
using a lattice fermion formulation which respects the chiral symmetry of continuum QCD. 
When
combined with the SM analysis of the experimental results for this
kaon decay channel, summarised as $|V_{us} f_+^{K\pi}(0)| =
0.2163(5)$ \cite{Antonelli:2010yf} (for $K^0 \to \pi^-$), it determines the
matrix element $|V_{us}|$. In combination with the other first-row
elements of the CKM matrix, $|V_{ub}|$ and $|V_{ud}|$, this allows
an immediate test for CKM unitarity in the SM and constrains models
for extensions of the SM. }

The history of recent efforts in predicting 
$\fpzero$~\cite{Bazavov:2013maa,Boyle:2007qe,Boyle:2010bh,Boyle:2013gsa,Bazavov:2012cd,Lubicz:2009ht}
is nicely summarised in~\cite{Colangelo:2010et,Aoki:2013ldr}.  
This quantity is currently known with an overall uncertainty of 0.3\% and
improving this precision further is mandatory in view of experimental
progress~\cite{AmelinoCamelia:2010me}.  
The error budget is typically dominated by the statistical uncertainty resulting 
from the  Monte Carlo sampling of the QCD path integral in lattice QCD.
Until recently the largest systematic uncertainty arose from the fact that 
simulations were only feasible for QCD with unphysically heavy pions. 
In order to make predictions for QCD as found in nature the simulation 
data had to be extrapolated using effective field theory or phenomenological 
models. 
Advances in 
algorithmic methods and computing technology now allow us to carry out 
simulations directly at the physical 
pion mass, thereby
removing the dominant systematic uncertainty from the extrapolation.
 
In this work we present the first prediction of the form factor $\fplus(0)$
with physical valence and sea quark masses in the continuum limit of 
domain wall lattice QCD with $N_f=2+1$ dynamical flavours.
The physics described by our simulations corresponds to nature 
up to isospin breaking in the light quark masses 
and electromagnetic effects and the contribution arising from charm (and
heavier quark) vacuum polarisation effects.

We anticipate the final results presented in this paper:
\begin{equation}
\fpzero=0.9685(34)(14)\,,\qquad |V_{us}|=0.2233(5)(9)\,,
\end{equation}
for the $K\to\pi$ form factor at vanishing momentum transfer and the 
CKM-matrix element for $u\to s$ flavour changing processes,
respectively.  The errors are statistical and systematic, respectively.

The rest of this paper is organised as follows: 
In Section~\ref{sec:Strategy} we explain the computational strategy for 
determining the form factor from Euclidean two- and three-point functions. 
In sections~\ref{sec:Simulation parameters} and~\ref{sec:Simulation results} 
we discuss  the simulation parameters and some aspects of 
the computational setup and the data analysis. Section~\ref{sec:Corrections} details
how we predict the physical results from a very small interpolation of the simulation data to the
precise physical quark mass combined with an extrapolation to the continuum limit.
A discussion of residual systematic errors follows in section~\ref{sec:Systematics}
and our final results and conclusions are presented in section~\ref{sec:Conclusions}.

\section{Strategy}\label{sec:Strategy}
In this section we define the observables from which we can
determine the $K\to\pi$ vector form factor $\fplus(q^2)$, where $q=p_K-p_\pi$
is the momentum transfer between the  kaon and pion.
The form factor is defined in terms of the QCD matrix element
\begin{equation} \label{eq:VME}
\langle \pi(p_\pi)|V_\mu|K(p_K)\rangle = 
	{ \fplus(q^2)} (p_K+p_\pi)_\mu+
	\fminus(q^2) (p_K-p_\pi)_\mu\,,
\end{equation}
of the flavour changing vector current $V_\mu=Z_V\, \bar u\gamma_\mu s$,
where $u$ and $s$ are up- and strange-quark fields and $Z_V$ is
the vector current renormalisation constant.
As first noted in~\cite{Na:2010uf} in the context of charm semileptonic 
decays an alternative way to determine the form factor is to consider
the matrix element of the flavour changing scalar current $S=\bar us$.
From the vector Ward-Takahashi identity we derive
\begin{equation}\label{eq:SME}
\langle \pi(p_\pi)|S|K(p_K)\rangle|_{q^2=0} = { f_0^{K\pi}(0)}\frac{m_K^2-m_\pi^2}{m_s-m_u}\,,
\end{equation}
which determines the vector form factor at zero momentum transfer
by means of the identity $\fpzero=f_0^{K\pi}(0)$.
Note that in the above equation the renormalisation 
constants of the scalar current and the quark masses cancel.

{In practice we determine the two matrix elements (\ref{eq:VME}) and (\ref{eq:SME}) from
the time dependence of combinations of Euclidean QCD two- and three-point
correlation functions which are the output of the actual simulation. 
The two-point function is defined as
\begin{equation}
C_i(t,\vec p_i) \equiv \sum_{\vec{x,y}} \langle
     \,O_{s_2,i}(t,\vec{y})\, O_{s_1,i}^\dagger(0,\vec{x})\,\rangle
   = \frac{Z_{s_1,i}Z_{s_2,i}^\ast}{2E_i} \left(e^{-E_it}+ e^{-E_i(T-t)}\right)\, ,
\label{eq:twopt}
\end{equation}
where $i=\pi$ or $K$, and $O_{s,i}$ are pseudoscalar interpolating
operators for the corresponding mesons, $O_{s,\pi}= \bar u\,\omega_s \gamma_5 d$\,
and $O_{s,K}=\bar d\,\omega_s \gamma_5 s$. The subscript $s$ indicates the
smearing type induced via the spacial smearing kernel $\omega_s$ which
in the simulations presented here is local ($s=L$, $\omega_L(\vec x,\vec y)=\delta_{\vec x,\vec y}$) 
or gauge-fixed wall ($s=W$, $\omega_W(\vec x,\vec y)=1$). In practice we employ $s_1=W$ and $s_2=L,W$.
The constants $Z_{s,i}$ are defined by 
$Z_{s,i}=\langle\,P_i\,|\,O_{s,i}^\dagger\,|\,0\,\rangle$ where 
$P_{i}$ is a pion or a kaon. As a result of our choice of boundary conditions 
$C_i(t,\vec p_i)$ is symmetric with respect to the middle of the temporal 
direction of length $T$ and
we have assumed that $t$ is such that the correlation
function is dominated by the lightest state (i.e. the groundstate 
pion or kaon) with 
 energy $E_i(\vec p_i)$. As explained further down meson momenta are induced 
using partially twisted boundary conditions. The three-point functions are defined as
(exposed here for the case $t_i=0$)
\begin{equation}
\begin{aligned}
C_{\Gamma,P_iP_f}(t_{i}=0,t,t_{f},\vec p_i,\vec p_f)
 &\equiv \sum_{\vec{x}_i,\vec{x},\vec x_f} 
    \langle\, O_{s_2,f}(t_{f},\vec x_f)\,
    \Gamma(t,\vec{x})\,O_{s_1,i}^\dagger(0,\vec x_i)\,\rangle\\
 &= \frac{Z_{s_2,i}\,Z_{s_1,f}}{4E_iE_f}\, \langle\,P_f(\vec{p}_f)\,|\,\Gamma\,|\,
        P_i(\vec{p}_i)\,\rangle\\
 &\hspace{-2.5cm}\phantom{=} \times\left\{\theta(t_f-t)\,e^{-E_i t-E_f(t_f-t)}\right.
   +c_\Gamma \left.\theta(t-t_f)\,e^{-E_i(T-t)-E_f(t-t_f)}\right\}\,,
\end{aligned}
\label{eq:3pt}
\end{equation}
where  $\Gamma\in \{V_\mu, \, S\}$ is
the vector/scalar current with flavour quantum numbers that allow for the 
$P_i\to P_f$ transition and 
 $Z_{s,f}=\langle\, 0\,|\,O_{s,f}|\,P_f\,\rangle$ where
again $P_f$ is either a pion or a kaon.
We will refer to $t_i$ ($t_f$) as the source (sink) time plane, respectively,
and $t$ is the time plane where the vector/scalar current is inserted.
The constant $c_\Gamma$ satisfies $c_{V_0}=-1$ for the time-component of the
vector current and $c_{V_i,S}=+1$
for the spatial components of the vector current and for the scalar current. 
In the second line of the above expression we have assumed that all time intervals are
sufficiently large for the lightest hadrons to give the dominant
contribution.
{Expressions~(\ref{eq:twopt}) and~(\ref{eq:3pt})
apply in slightly modified form in our earlier 
simulations~\cite{Boyle:2007qe,Boyle:2010bh,Boyle:2013gsa} due
to differences in technical details 
(noise- and sequential source propagators~\cite{Boyle:2008rh}).
}
}

We  consider two strategies for determining the form factor $\fpzero$ from the above 
correlation functions. The first  consists of a simultaneous fit over all
two- and three-point functions from which the matrix elements~(\ref{eq:VME})  and~(\ref{eq:SME})
are readily extracted. The other strategy is to 
determine both matrix elements from fits to 
suitably chosen ratios of correlation 
functions~\cite{Boyle:2007wg}, e.g.,
\begin{equation}\label{eq:ratio}
 \begin{array}{l}
\displaystyle
 R_{\Gamma,\,K \pi }(t,\vec{p}_K,\vec{p}_\pi)=
 4\sqrt{{E_K  E_\pi  }}\sqrt{
 \frac{C_{\Gamma,K \pi }(0,t,t_{\rm snk},\vec p_K ,\vec p_\pi)
        \,C_{\Gamma,\pi K }(0,t,t_{\rm snk},\vec p_\pi ,\vec p_K)}
 {\tilde C_{K}(t_{\rm snk},\vec p_K)\,
    \tilde C_{\pi}(t_{\rm snk},\vec p_\pi)}}\\[6mm]
 \hspace{27mm}\stackrel{0\ll t\ll t_{snk}}{=}
 \langle \pi(p_\pi)|\Gamma|K(p_K) \rangle\,.
 \end{array}
\end{equation}
In the denominator we introduced the function $\tilde C_{K/\pi}(t,\vec p)=
C_{K/\pi}(t,\vec p)- \frac 12 C_{K/\pi}(T/2,\vec p)\,e^{E_{K/\pi} (T/2-t)}$
for $t<T/2$
to subtract the around-the-world contribution due to the periodic 
boundary condition the mesons experience in the time-direction.

We obtain the vector current renormalisation constant $Z_V$ by imposing conservation
of the electromagnetic charge, i.e. $f^{\pi\pi/KK}(0)=1$ and hence,
\begin{equation}
\label{eq:zv}
Z_V^{\pi,K} = \frac{\tilde C_{\pi,K}(t_f,\vec 0)}
    {C^{(B)}_{V_0,\pi\pi,KK}(t_i,t,t_f,\vec{0},\vec{0}\,) }\,.
\end{equation}
The superscript $B$ in the denominator indicates
that we take the bare (unrenormalised) current in the three-point
function. While both $Z_V^\pi$ and $Z_V^K$ obtained in this
way renormalise the flavour-changing vector
current in~(\ref{eq:VME}) we note that they 
differ by mass dependent cutoff effects. We will return
to this point in later sections when discussing the 
continuum limit.\\

As a result of the quantisation of quark- and hadron-momenta in a finite 
lattice box the desired kinematical situation $q^2=0$ 
is generally not directly accessible.  
We follow~\cite{Bedaque:2004ax,deDivitiis:2004kq,Sachrajda:2004mi,
Boyle:2007wg,Boyle:2010bh} and use partially twisted boundary conditions for
the quark fields, i.e. $\psi(x+L \hat i)=e^{i\theta_i/L}\psi(x)$, where $L$ is the
spatial extent of the lattice, $i$ one of the spatial directions and $\psi$ 
one of the up- or strange-quark fields. Varying the twist angle we expect
the meson energies to follow the dispersion relation 
\begin{equation}
E^2(\vec p^2) = m^2 +{\vec p}^2\,,
\label{eq:pi_disprel}
\end{equation}
where ${\vec p}L\equiv\Delta \vec \theta$ is the difference of twist angles imposed on the
valence quark- and antiquark of the respective meson. 
The kinematical point $q^2=0$ can be reached in all our simulations by suitably tuning
the quark boundary conditions~\cite{Boyle:2007wg}. In particular, enforcing
\begin{equation}\label{eq:mom_transfer}
0\stackrel{!}{=}q^2=(p_K-p_\pi)^2=\big(E_K(\vec p_K)-E_\pi(\vec p_\pi)\big)^2
        -\big(\vec{p}_K -\vec{p}_\pi\big)^2\,,
\end{equation}
we make use of two specific choices for the twist angles~\cite{Boyle:2007wg},
\begin{equation}\label{eq:twists}
 \begin{array}{llcccc}
&
  |\vec{\theta}_K| =
           L\sqrt{({m_K^2+m_\pi^2 \over 2m_\pi})^2 - m_K^2}
      &\textrm{and}&\vec{\theta}_\pi=\vec{0}\,,\\[2mm]
{\rm or}&
          |\vec{\theta}_\pi| =L
          \sqrt{({m_K^2+m_\pi^2 \over 2m_K})^2 -
          m_\pi^2}&\textrm{and}
      &\vec{\theta}_K =\vec{0}\,.\\[2mm]
 \end{array}
\end{equation}

As is evident from eqn.~(\ref{eq:SME}) the vector form factor at zero momentum 
transfer can  be extracted directly from a fit to the three point functions 
of the scalar current. The vector current matrix element in eqn.~(\ref{eq:VME})
is instead parameterised in terms of two form factors. These are readily extracted from
a simultaneous fit to the correlation function data for both time- and space-components 
of the vector current.
\section{Simulation parameters}\label{sec:Simulation parameters}
\begin{table*}
\begin{center}
        \small
\begin{tabular}{l@{\hspace{3mm}}l@{\hspace{2mm}}c@{\hspace{2mm}}c@{\hspace{2mm}}c@{\hspace{2mm}}c@{\hspace{2mm}}cl@{\hspace{2mm}}l@{\hspace{2mm}}l@{\hspace{2mm}}l}
 \hline\hline &&&&&\\[-3.0ex]
 set &$\beta$ & $a\,/\mathrm{fm}$ & $L/a$ & $T/a$ & $am_q$ & $am_s^\mathrm{sea}$ & $am_s^{\rm val}$&  $\frac{m_\pi}{\rm MeV}$&$m_\pi L$\\[.5mm]
 \hline
 A$_3$   &2.13 &0.11 &24 &64  &0.0300&0.040&0.040&693 &9.3 \\
 A$_2$   &2.13 &0.11 &24 &64  &0.0200&0.040&0.040&575 &7.7 \\
 A$_1$   &2.13 &0.11 &24 &64  &0.0100&0.040&0.040&431 &5.8 \\
 A$_5^4$ &2.13 &0.11 &24 &64  &0.0050&0.040&0.040&341 &4.6 \\
 A$_5^3$ &2.13 &0.11 &24 &64  &0.0050&0.040&0.030&341 &4.6\\
 C$_8$   &2.25 &0.08 &32 &64  &0.0080&0.030&0.025&431 &5.5\\
 C$_6$   &2.25 &0.08 &32 &64  &0.0060&0.030&0.025&360 &4.8\\        
 C$_4$   &2.25 &0.08 &32 &64  &0.0040&0.030&0.025&304 &4.1\\       
 A$_{\rm phys}$ &2.13 &0.11 &48&96 &0.00078 &0.0362 &0.0362 &139 &3.8\\
 C$_{\rm phys}$ &2.25 &0.08 &64&128&0.000678&0.02661&0.02661&139 &3.9\\
 \hline\hline
\end{tabular}
\end{center}
\caption{Basic parameters for all ensembles of gauge field configurations.}
\label{tab:ensemble-params}
\end{table*}

This report centres around two new ensembles with physical light quark masses in 
large volume~\cite{RBCUKQCDPhysicalPoint}.  
The data analysis will however also take advantage of the information provided
by ensembles with unphysically heavy pions which we generated and discussed
earlier~\cite{Allton:2008pn,Aoki:2010pe,Aoki:2010dy}.

The gauge field ensembles represent QCD with $N_f=2+1$ 
dynamical flavours at two different lattice spacings. The basic parameters of these
ensembles are listed in Table~\ref{tab:ensemble-params}.
All ensembles have been generated with the 
Iwasaki gauge action~\cite{Iwasaki:1984cj,Iwasaki:1985we}.
For the discretisation of the quark fields 
we adopt the domain wall fermion (DWF) action with the M\"obius 
kernel~\cite{Brower:2004xi,Brower:2005qw,Brower:2012vk} 
for the ensembles with physical quark mass and otherwise Shamir 
DWF~\cite{Kaplan:1992bt,Shamir:1993zy}.
The difference between both kernels in our implementation
corresponds to a rescaling such that M\"obius domain wall fermions
are equivalent to Shamir domain wall fermions at twice the extension in the fifth 
dimension~\cite{RBCUKQCDPhysicalPoint}. 
M\"obius domain wall fermions are hence cheaper to simulate while providing
the same level of lattice chiral symmetry. Results from both formulations of domain wall
fermions lie on the same scaling trajectory towards the continuum limit
with cutoff-effects starting at $O(a^2)$. 
Even these $O(a^2)$ cutoff-effects themselves are 
expected to agree between our Mobius and Shamir 
formulations, with their relative difference at or 
below the level of 1\%~\cite{RBCUKQCDPhysicalPoint}.

We have used eqn.~(\ref{eq:twists}) to determine our choice of twist angles. The
values for the unphysical ensembles are given in~\cite{Boyle:2013gsa}. 
In our previous studies of the  $K\to\pi$ form factor
we found that the kinematical situation with the kaon at rest
provides for a  better signal-to-noise ratio. On
the physical point ensembles we therefore only twist the up quark in the pion 
while leaving the kaon at rest. Our choice is
$\vec\theta_\pi^T=\frac {2\pi}L(0.5893,0.5893,0.5893)$ on ${\rm A_{\rm phys}}$ and
$\vec\theta_\pi^T=\frac {2\pi}L(0.5824,0.5824,0.5824)$ on ${\rm C_{\rm phys}}$ with
$\vec \theta_K^T=(0,0,0)$ in each case. 

\section{Simulation results}\label{sec:Simulation results}
In this section we present  all steps involved in the
analysis towards the determination
of the form factor on the physical point ensembles. Apart from details which 
we mention in the text, the analysis for the unphysical-point 
ensembles follows~\cite{Boyle:2007qe,Boyle:2010bh,Boyle:2013gsa}.

\subsection{Correlation functions and AMA}
{
The substantial cost of solving for light quark propagators on the large volume, physical pion mass $A_{\rm phys}$ and $C_{\rm phys}$ ensembles required us to make several algorithmic refinements to our measurement strategy. All correlation functions associated with these ensembles were computed using Coulomb gauge-fixed wall source propagators, together with the all-mode averaging (AMA) technique introduced in \cite{Blum:2012uh}. In the AMA formalism one replaces a direct calculation of an expensive lattice observable $\mathcal{O}$ with a less-expensive approximation $\mathcal{O}'$ and a correction term $\Delta \mathcal{O}$. The lattice action and ensemble averages $\langle \mathcal{O} \rangle$, $\langle \mathcal{O}' \rangle$, and $\langle \Delta \mathcal{O} \rangle$ are all assumed to be invariant under a group $G$ of lattice symmetries. We define the AMA estimator by averaging the inexpensive approximation $\mathcal{O}'$ over some number $N$ of transformations $g \in G$, and applying the correction term $\Delta O$:
\begin{equation}
\label{eqn:AMA}
\mathcal{O}_{\rm AMA} = \frac{1}{N} \sum_{g \in G} \mathcal{O}_{g}' + \Delta \mathcal{O},
\end{equation}
where the notation $\mathcal{O}_{g}$ denotes $\mathcal{O}$ computed after $g$ is applied. We find in practice that the statistical error per unit of computer time can be markedly reduced using AMA with a judicious choice of $\mathcal{O}'$ and $\Delta \mathcal{O}$, relative to computing $\mathcal{O}$ directly.
}

{
In the context of this calculation the relevant lattice symmetry is the group of translations in the temporal direction. Quark propagators were computed using a deflated mixed-precision conjugate gradient (CG) solver, with 600 (1500) single-precision low-mode deflation vectors obtained from the EigCG algorithm applied to a four dimensional volume source on the $A_{\rm phys}$ ($C_{\rm phys}$) ensemble. We further distinguish between \textit{exact} and \textit{sloppy} light quark propagators. Exact light quark propagators were computed using a tight CG stopping residual $r = 10^{-8}$ for 7 (8) time slices. To avoid bias associated with the even-odd preconditioning used in the CG solves we randomly shifted the time slices used to compute exact propagators on each configuration. Sloppy light quark propagators were computed using a reduced precision $r = 10^{-4}$ and for all time slices. Strange quark propagators were sufficiently inexpensive that exact solves were computed for all time slices. For a given two- or three-point function we then constructed a sloppy estimate ($\mathcal{O}'$) for all time slices with the sloppy light quark propagators, and a correction term ($\Delta \mathcal{O}$) using the exact light quark propagators on time slices for which these are available. We then compute the AMA estimator according to \eqref{eqn:AMA}, after averaging $\mathcal{O}'$ over all time translations. The full measurement package, which also computes observables related to the $K \rightarrow (\pi \pi)_{I=2}$ decay \cite{Blum:2015ywa} and other low-energy QCD observables \cite{RBCUKQCDPhysicalPoint} from the same propagators, took 5.5 days per configuration on the $A_{\rm phys}$ ensemble using 1 rack (1024 nodes) of IBM Blue Gene/Q hardware, and 5.3 hours per measurement on the $C_{\rm phys}$ ensemble using 32 racks of Blue Gene/Q sustaining 1.2PFlop/s. Additional details of the calculation can be found in \cite{RBCUKQCDPhysicalPoint}.
}

The above set of quark propagators allows generating 
AMA three-point functions where at least one of the 
quarks coupling to the external current is a strange quark,
for all possible source-sink-separations $\Delta t=|t_f-t_i|$ up to $T/2a$ (e.g. $K\to\pi$ and $K\to K$);
in each case based on $T/a$ different positions of the source plane. Results at 
constant $\Delta t$ from different source planes were averaged into one bin on every
configuration.
For our choice of the 7 (8) source planes for the exact light 
quark solves on the $A_{\rm phys}$ ($C_{\rm phys}$) ensembles the $\pi\to\pi$ three-point 
function entering e.g. the determination of $Z_V^\pi$ from eqn.~\eqref{eq:zv} can be computed on  
every fourth (fifth) source-sink-separation following the AMA prescription.

{In all cases we use the bootstrap resampling technique with 500 samples to determine the statistical errors.}
\subsection{Extracting the form factor}
{
We consider two ways of extracting the form factors: 
\bi
 \item[a)] simultaneously fit
\bi
       \item the pion and kaon two-point functions 
	--- $C_{\pi,K}(t,\mathbf{p})$,
       \item  the three-point functions determining $Z_{V}^{\pi}$ and $Z_{V}^{K}$ --- 
        $C_{V_{0},\pi \pi,K K}^{(B)}(t_{i},t,t_{f},\mathbf{0},\mathbf{0})$,
       \item  the three-point functions determining  $f_{+}^{K \pi}(0)$ through the 
	vector and scalar current, as well as their time-reversed counterparts --- 
	$C_{\Gamma,K \pi}(t,t_{\mathrm{snk}},\mathbf{p}_{K},\mathbf{p}_{\pi})$ and 
	$C_{\Gamma,\pi K}(t,t_{\mathrm{snk}},\mathbf{p}_{K},\mathbf{p}_{\pi})$, 
\ei
	by minimising a single, global $\chi^{2}$.
	Two-point functions are fit to the right-hand side of eqn.~(\ref{eq:twopt}), 
	and three-point functions are fit to the right-hand side of eqn.~(\ref{eq:3pt}). 
\item[b)] determine the result for the ratios $R_{V_0,K\pi}^{}$,
	$R_{V_i,K\pi}$ and $R_{S,K\pi}$ by taking their value 
	at $t=\Delta t/2$ for even $\Delta t$ and the weighted average of the two 
	time-slices adjacent to $\Delta t/2$ for odd $\Delta t$. 
	In this way we
	avoid having to choose a suitable \emph{plateau region} subjectively by hand.
	With the values for the ratios at hand we can solve~(\ref{eq:VME}) 
	and~(\ref{eq:SME}) for the vector form factor.
	In the next section we describe how remaining contaminations by
	excited states are dealt with.
\ei
We find that  the values for the form factor obtained from the two analyses agree.
All results presented in the following are based on method (b) for the following reasons:
we observe as much 
as a factor of 5 difference in the statistical error on $Z_{V}^{\pi,K}$ between 
the ratio fit approach and the global fit approach; for this particular 
quantity the ratio (\ref{eq:zv}) is clearly superior  since the measurement of $Z_V^{\pi,K}$ 
is not contaminated by excited states and hence the operator can be placed closer to the source/sink 
leading to reduced statistical errors. We also observe that 
the three-point functions $C_{\Gamma,K \pi}$ and $C_{\Gamma,\pi K}$ are not 
symmetric between the source and sink walls, since the initial and final states 
are different, making it a priori difficult to decide on sensible fit 
ranges for extracting the form factors. 
}
\subsection{Excited state contamination}
The availability  of data for a large
range of source-sink separations in the three-point functions allows us to 
study excited states in detail and to reduce their contribution
to a minimum. 
\begin{figure}
	\begin{center} 
		\includegraphics[width=7.5cm]{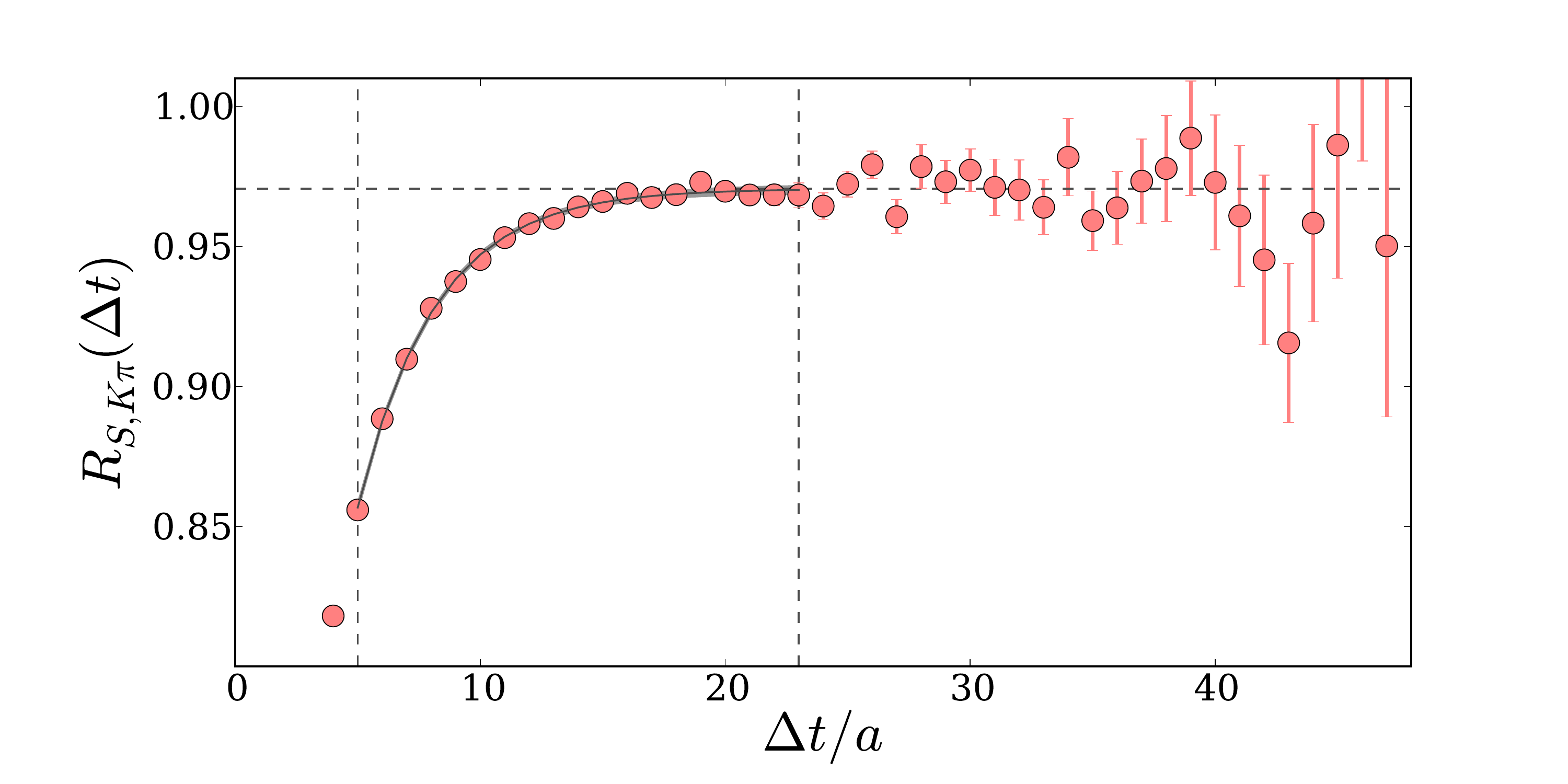}\hspace{-5mm}
		\includegraphics[width=7.5cm]{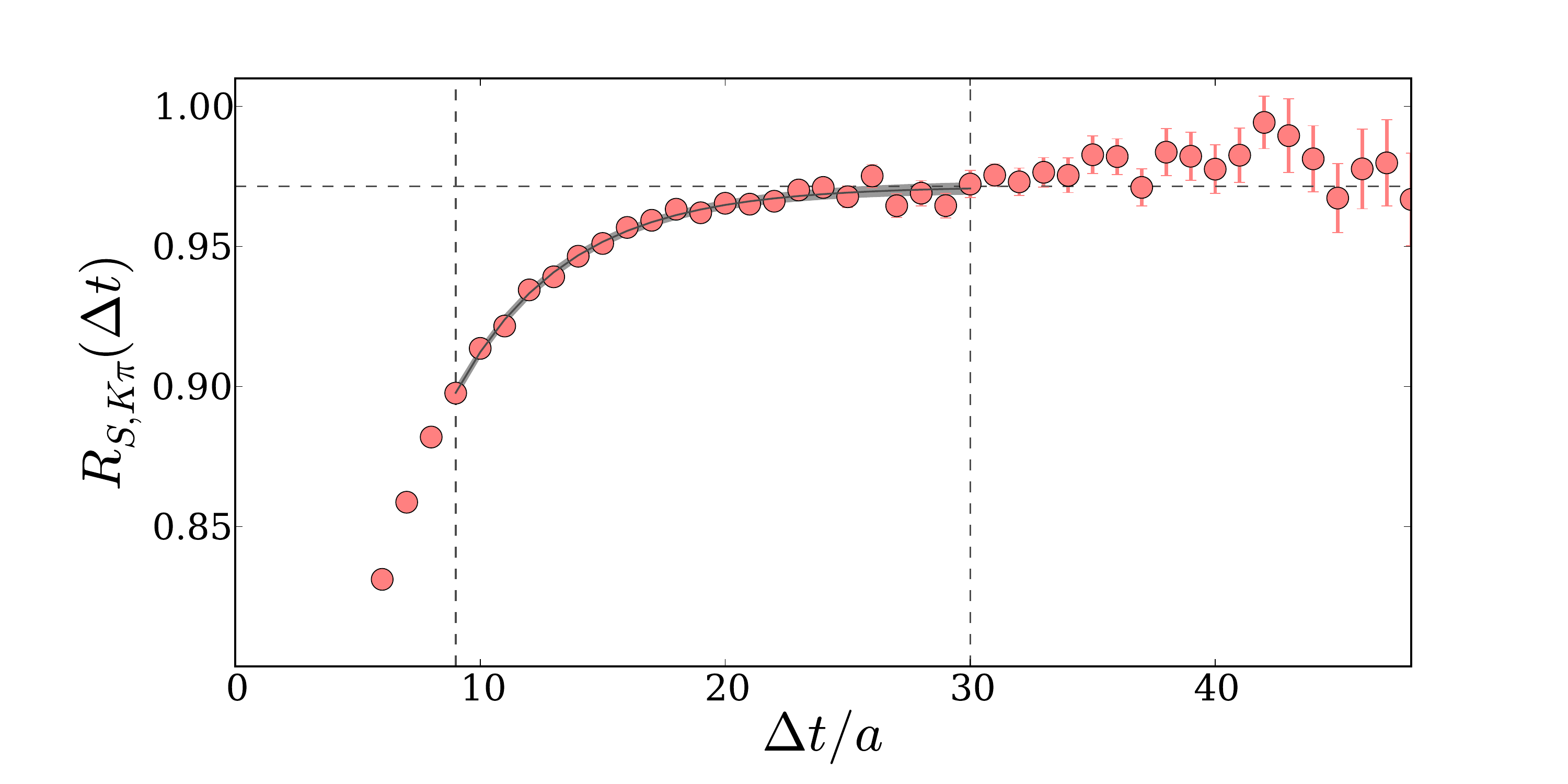}
		\includegraphics[width=7.5cm]{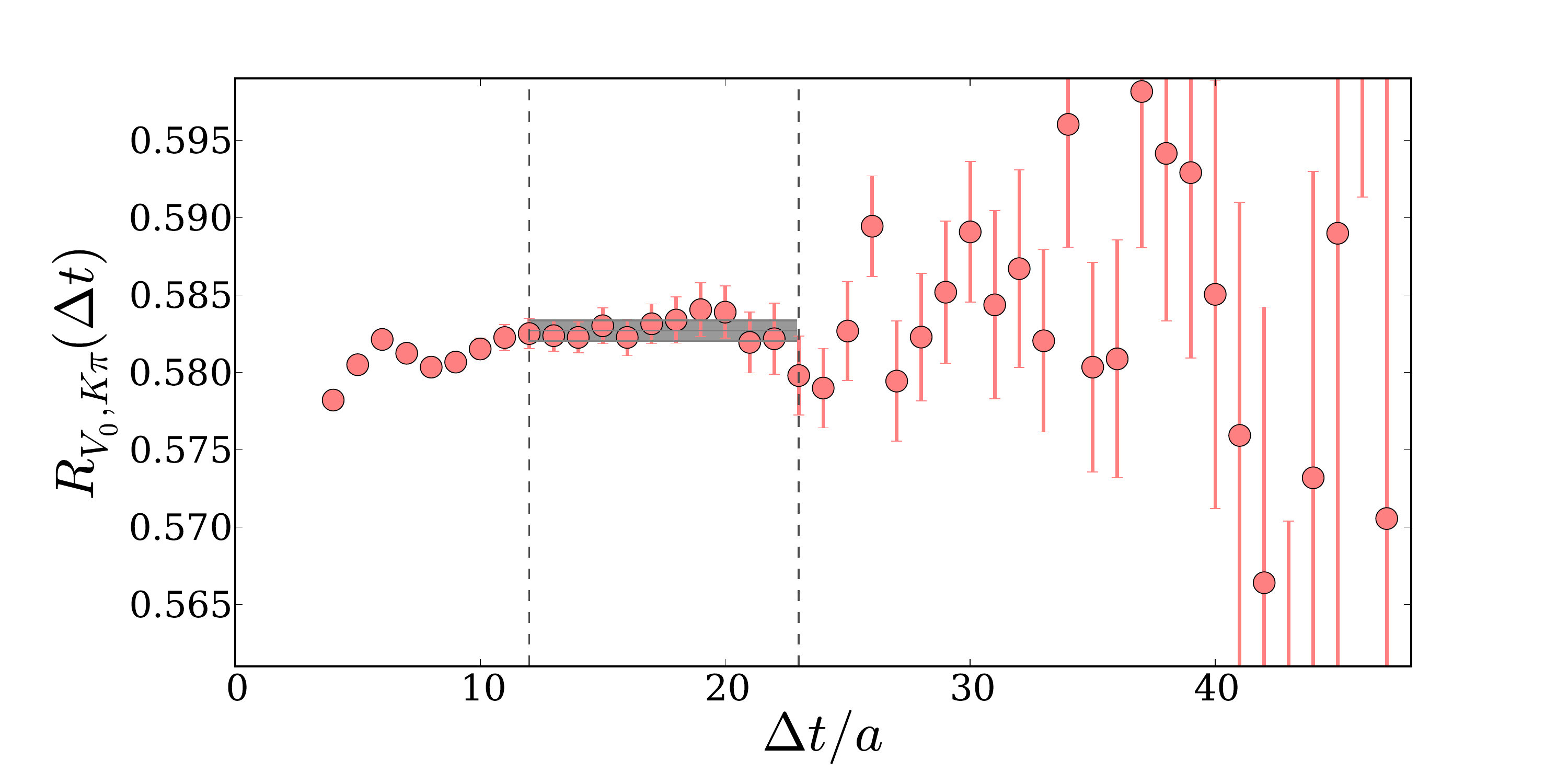}\hspace{-5mm}
		\includegraphics[width=7.5cm]{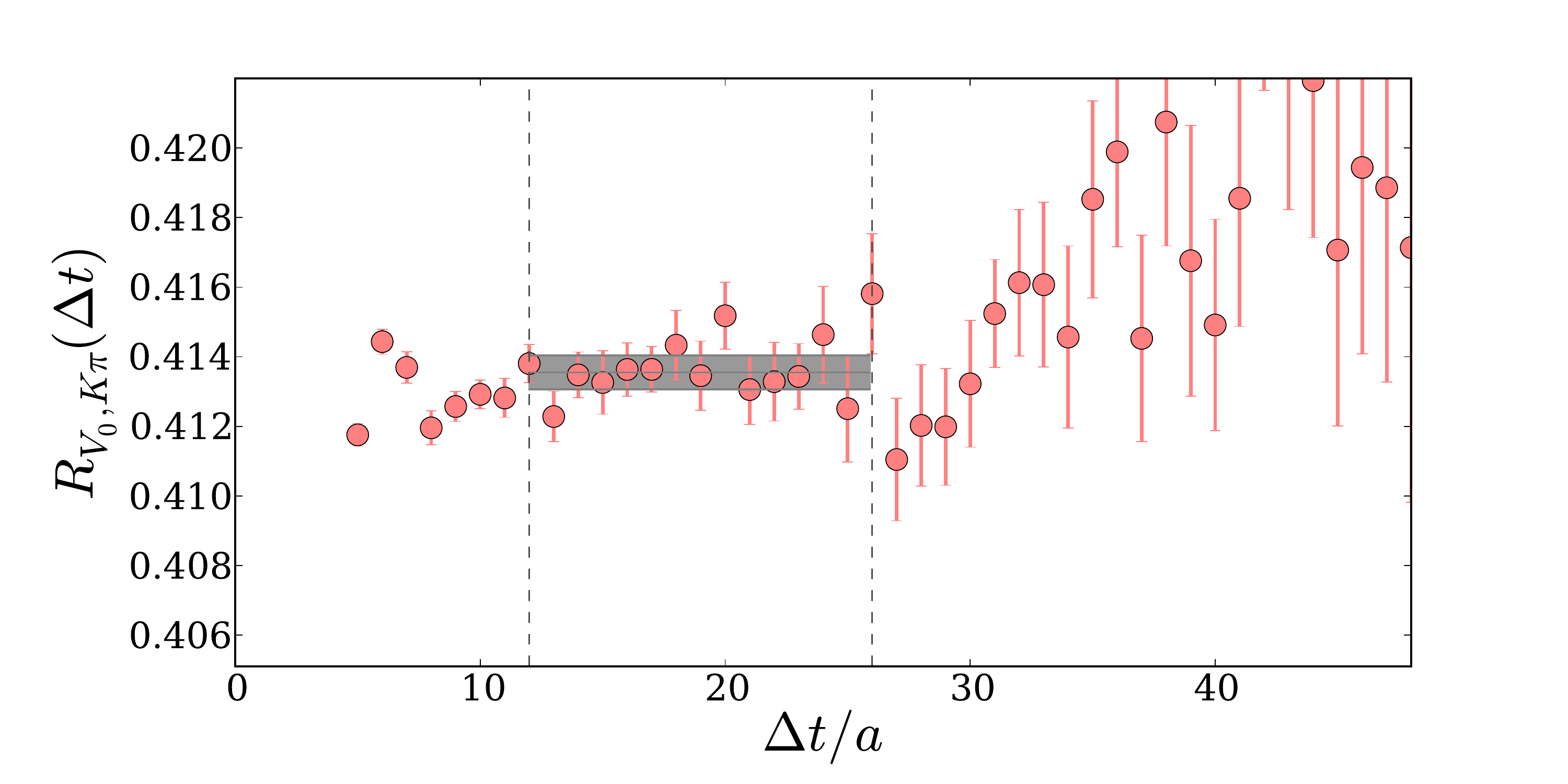}
		\includegraphics[width=7.5cm]{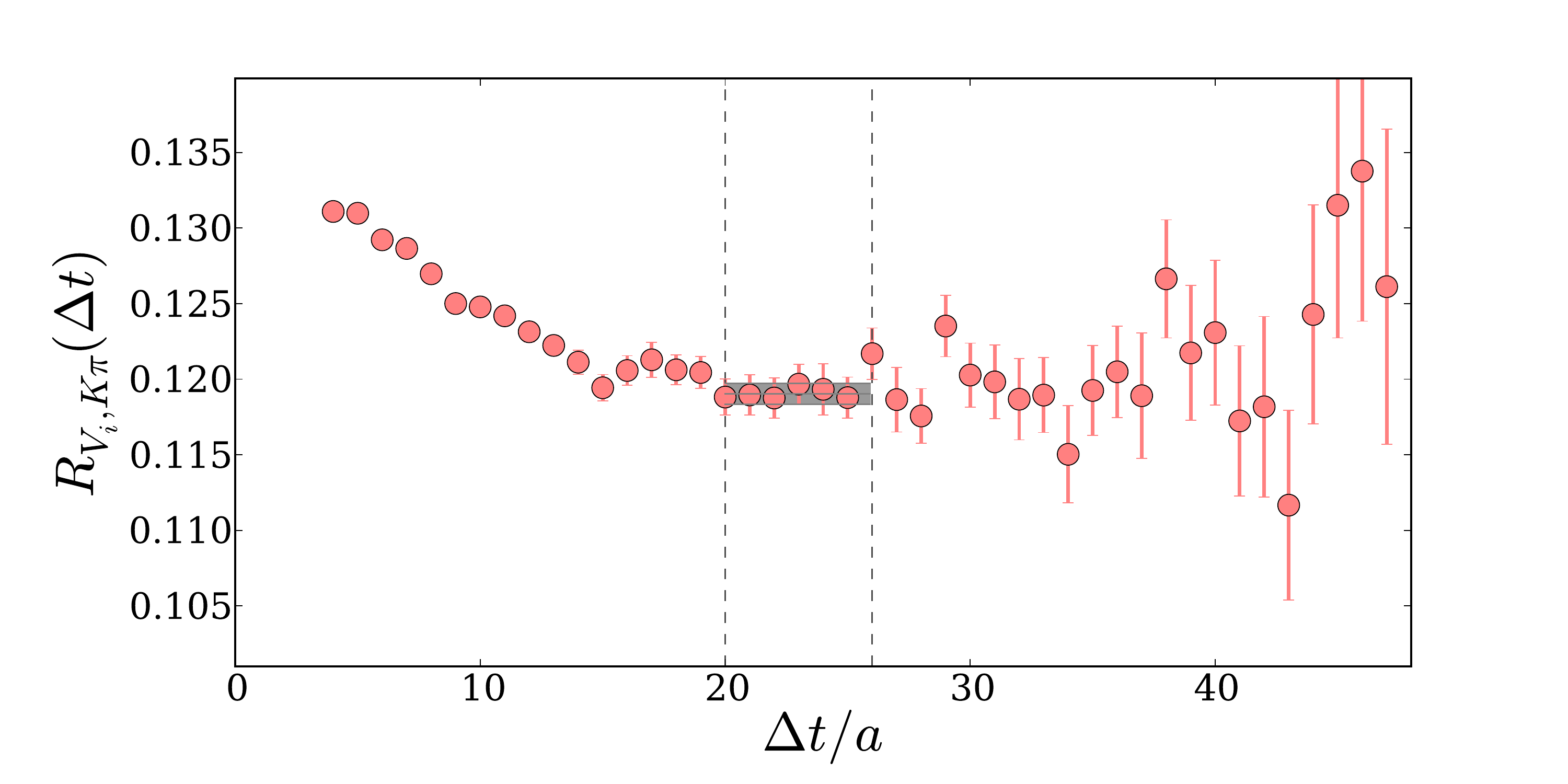}\hspace{-5mm}
		\includegraphics[width=7.5cm]{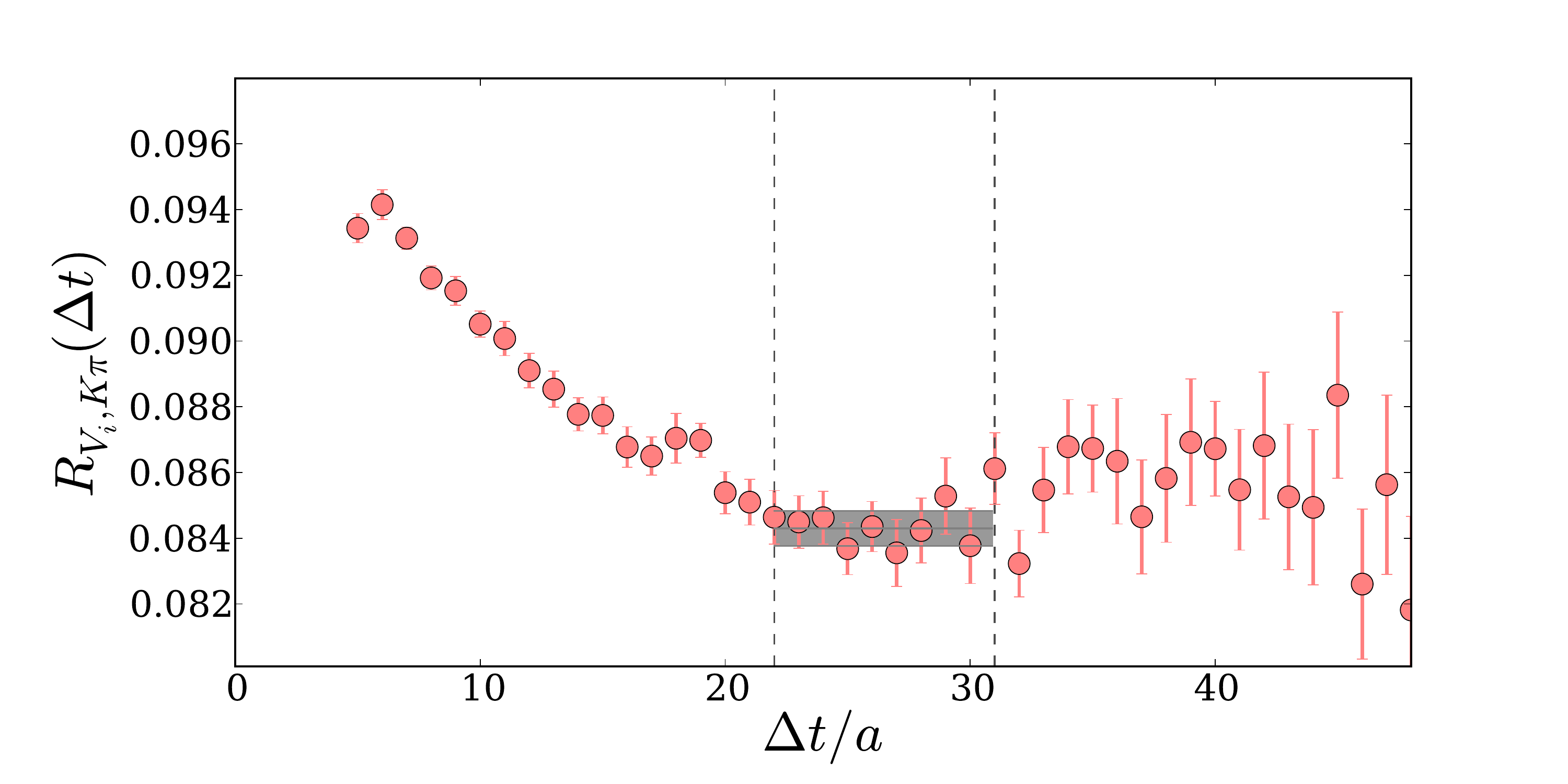}
 	\end{center}
	\caption{Dependence of ratios $R_{S/V_0/V_i,K\pi}$ on the source-sink separation $\Delta t$.
	Vertical lines indicate the choice of fit range. The fit result is indicated by
	the shaded area which in the case of the scalar current (first row of plots)
	results from an exponential parameterisation. In this case we also indicate the
	result for the constant term as dashed horizontal line.}\label{fig:fits}
\end{figure}
Figure~\ref{fig:fits} shows a non-trivial dependence on the source-sink separation
$\Delta t$ of $R_{V_0,K\pi}^{}$,
$R_{V_i,K\pi}$ and $R_{S,K\pi}$ as defined in eqn.~\eqref{eq:ratio}.
$R_{V_0,K\pi}$ is stable starting from surprisingly small values
for $\Delta t$. $R_{V_i,K\pi}$ 
and $R_{S,K\pi}$ on the other hand 
show a rather noticeable dependence on the source-sink separation. 
The dependence of $R_{S,K\pi}$ on $\Delta t$ can be parameterised very well with a
single exponential term added to the constant term which we expect to dominate 
for large $\Delta t$. This procedure
is stable under variation of the smallest and largest values for $\Delta t$ included in the fit.
While this provides for a good way of determining $\fpzero$ an alternative approach 
is to determine from 
the exponential fit the smallest value of $\Delta t$ where excited state contributions are
well below the statistical error. Fitting a constant to $R_{S,K\pi}$ above this minimum 
value for $\Delta t$ results in a compatible central value and approximately the same
statistical error. 
The latter fit shows more sensitivity to the range of $\Delta t$ included
in the fit and we therefore decide to take the result from the exponential parameterisation.

The determination of $\fpzero$ from $R_{V_{0,i},K\pi}$ proceeds differently since the corresponding
ground state matrix element is parameterised in terms of two form factors. In order
to take advantage of the data from the large number of source-sink separations at hand
we minimise the corresponding global $\chi^2$-function.

The contamination at small $\Delta t$ seen in particular in  $R_{V_{i},K\pi}$ could in principle be
parameterised as above by adding an exponential. This however turned out to 
result in unstable $\chi^2$-minimisations. Whilst the $\Delta t$ dependence in 
$R_{S,K\pi}$ was sufficiently pronounced for stable fit results, the combination 
of a milder $\Delta t$ dependence together with a worse signal to noise ratio
did not allow us to proceed in this way for $R_{V_i,K\pi}$. 
Instead we vary the minimum and maximum values of $\Delta t$ entering the above 
function minimisation for both $R_{V_{0},K\pi}$ and $R_{V_{i},K\pi}$  over a wide range. We 
were able to find a stable region from which we determine the final results. We illustrate
the fit ranges and 
fit results as shaded  bands in the second and third row of plots in figure~\ref{fig:fits}.

The determination of the renormalisation constants $Z_V^{\pi,K}$ proceeds along very similar lines.  
There is very little dependence of the fit result on $\Delta t$ and we determine it by fitting 
a constant.
\begin{table}
\begin{center}
 \footnotesize
\begin{tabular}{l@{\hspace{1mm}}|c@{\hspace{1mm}}c|l@{\hspace{1mm}}l|l@{\hspace{1mm}}l@{\hspace{1mm}}l@{\hspace{1mm}}l}
 \hline\hline
        &         &       &&&\multicolumn{3}{c}{$\fpzero$}\\
 set 	&$am_\pi$ &$am_K$ &\multicolumn{1}{c}{$Z_V^\pi$} &
                           \multicolumn{1}{c|}{  $Z_V^K$}&
			   \multicolumn{1}{c}{$Z_V^\pi V$} &
		           \multicolumn{1}{c}{  $Z_V^K V$}&
			   \multicolumn{1}{c}{$S$} \\[-0mm]
 \hline&&&&&\\[-4mm]
 A$_3$   	&0.38840(39) & 0.41628(39) & 0.716106(77) & 0.717358(75) & 0.998289(79) & 1.000033(80) &  \\
 A$_2$   	&0.32231(47) & 0.38438(46) & 0.71499(12) & 0.717252(93) & 0.99404(29) & 0.99719(28)    &  \\
 A$_1$   	&0.24157(38) & 0.35009(39) & 0.71408(20) & 0.717047(74) & 0.98474(89) & 0.98884(90)    &  \\
 A$_5^4$ 	&0.19093(46) & 0.33197(58) & 0.71399(58) & 0.71679(13) & 0.9746(43) & 0.9784(43) & 0.9793(46) \\
 A$_5^3$ 	&0.19093(45) & 0.29818(52) & 0.71399(58) & 0.71570(16) & 0.9850(27) & 0.9874(27) & 0.9878(47) \\
 C$_8$		&0.17249(50) & 0.24125(47) & 0.74435(40) & 0.74580(12) & 0.9890(17) & 0.9909(17) &  \\
 C$_6$   	&0.15104(41) & 0.23276(45) & 0.74387(56) & 0.74563(13) & 0.9833(24) & 0.9857(24) & 0.9796(39) \\
 C$_4$   	&0.12775(41) & 0.22624(51) & 0.74480(94) & 0.74585(16) & 0.9805(39) & 0.9819(35) & 0.9796(47) \\
 A$_{\rm phys}$ &0.08046(11) & 0.28856(14) & 0.71081(14) & 0.714051(20) & 0.9703(16) & 0.9747(16) & 0.9712(14) \\
 C$_{\rm phys}$	&0.059010(95) & 0.21524(11) & 0.742966(81) & 0.745121(23) & 0.9673(18) & 0.9701(17) & 0.9707(21) \\
 \hline\hline
\end{tabular}
\end{center}
\caption{Simulation results that enter the data analysis. For the form factor we consider the results
	obtained from the vector current matrix element renormalised with $Z_V^\pi$ and $Z_V^K$, 
	respectively, and we also consider the result determined from the scalar current matrix element. }\label{tab:ffresults}
\end{table}

\begin{figure}
	\begin{center}
		\includegraphics[width=7.0cm]{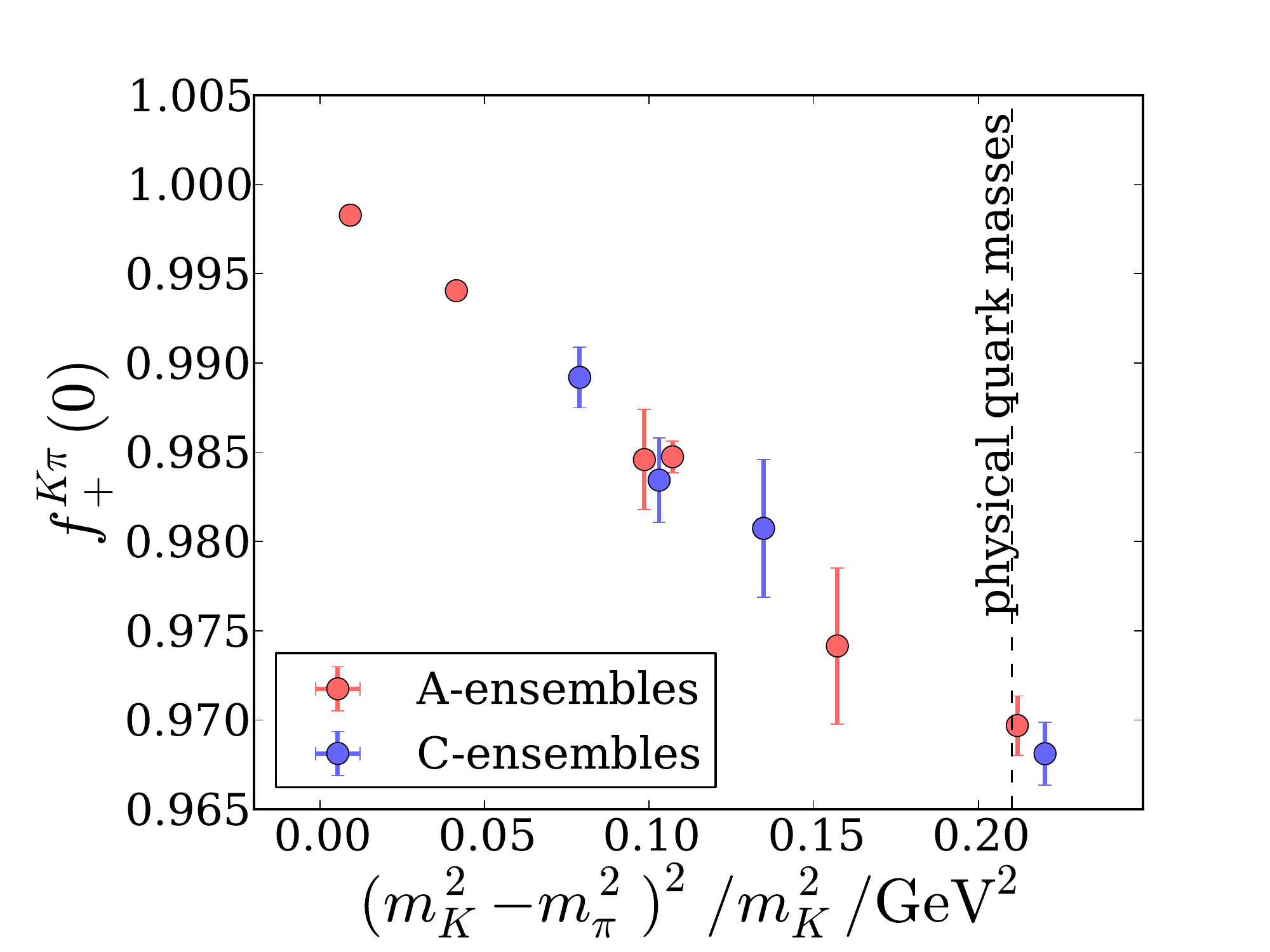}\hspace{-0mm}
		\includegraphics[width=7.0cm]{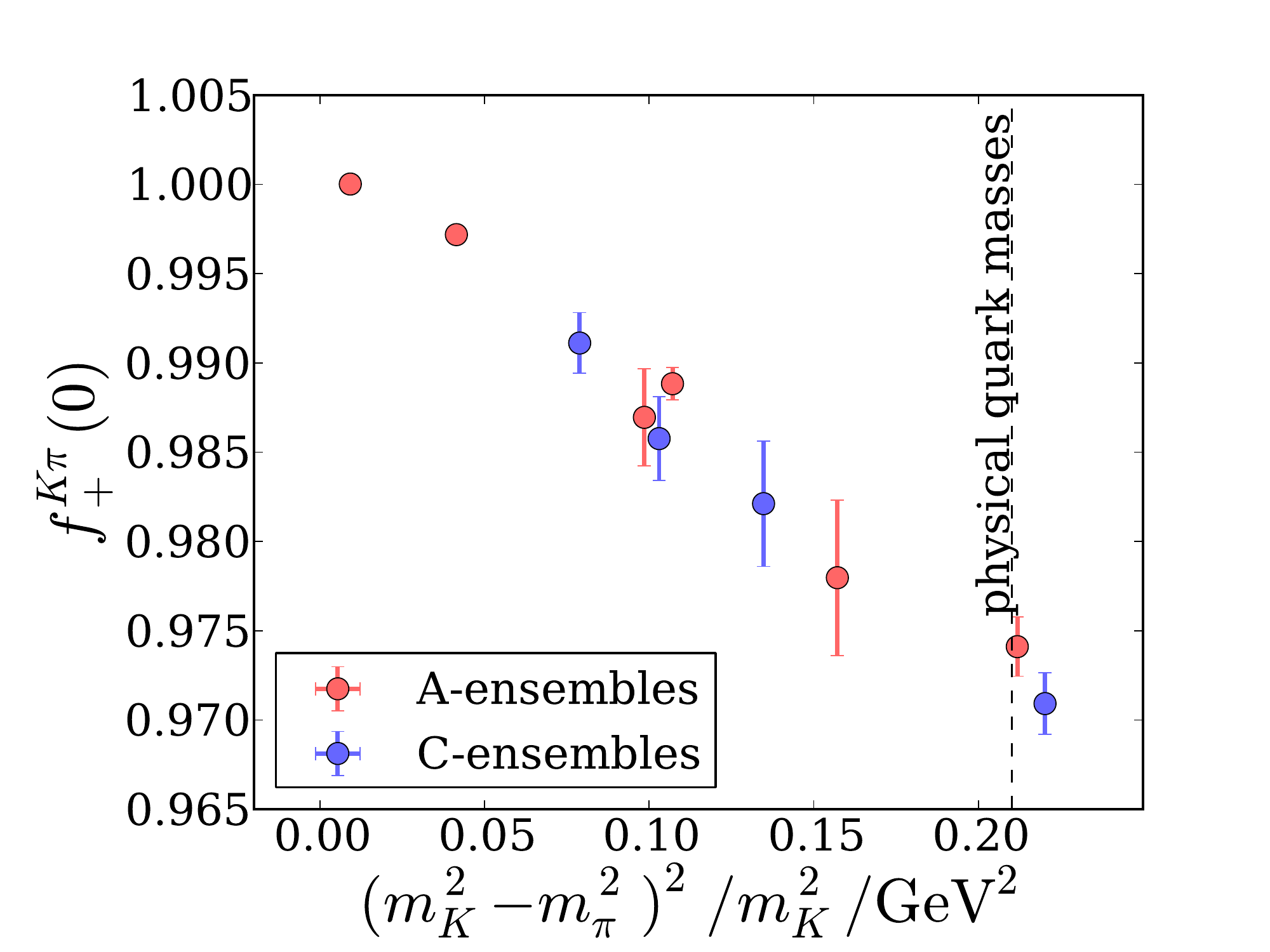}\\
		\includegraphics[width=7.0cm]{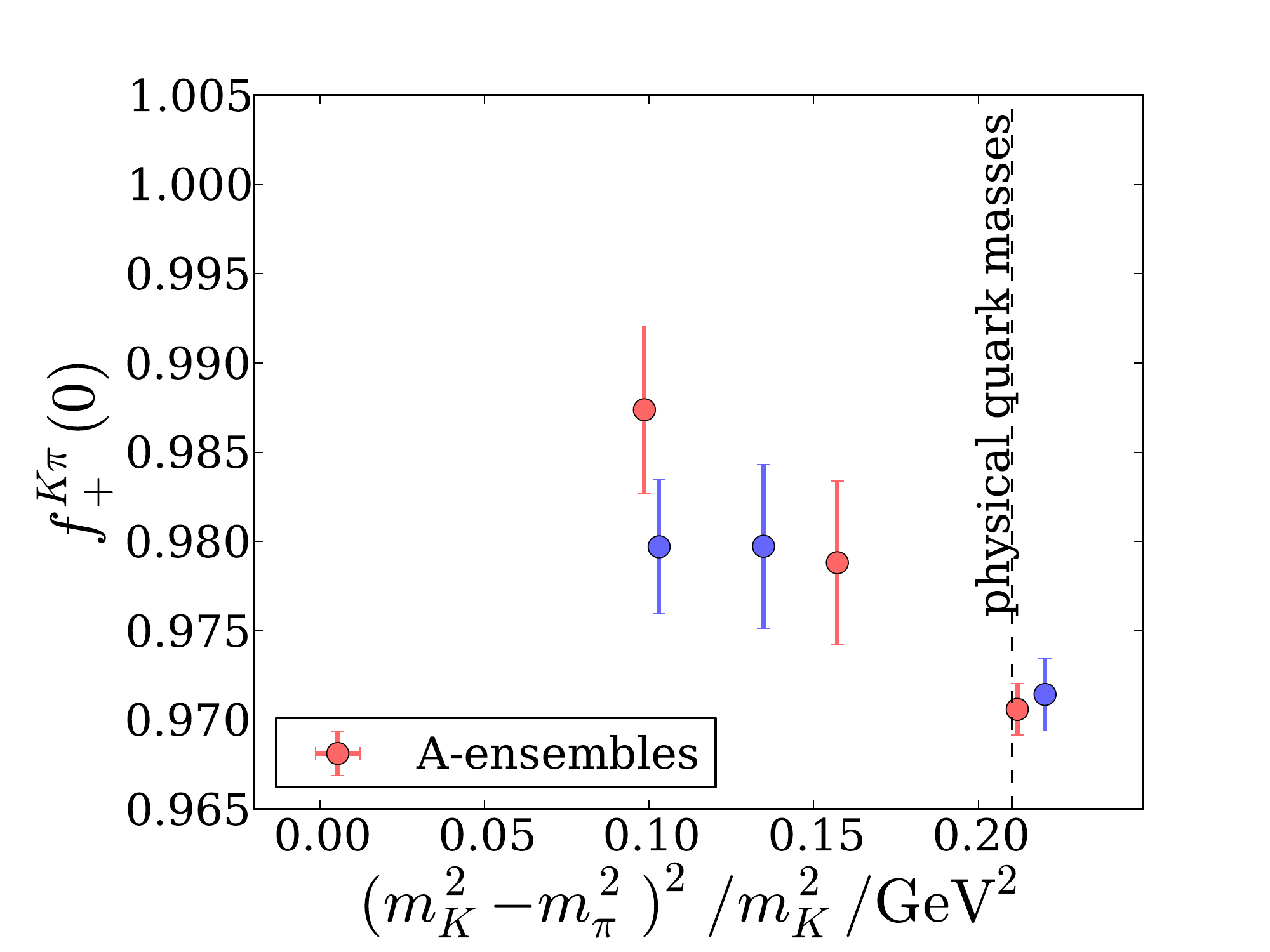}
 	\end{center}
	\caption{Summary of all simulation data for $\fpzero$ from the vector current
		renormalised with $Z_V^\pi$ (left), from the vector current renormalised with $Z_V^K$ (middle)
		 and from the scalar current (right).}\label{fig:final}
\end{figure}
\subsection{Simulation results for the form factor}\label{sec:ffres}
The valence quark boundary conditions in our simulations were determined using 
equations~(\ref{eq:twists}) with estimates for the
pion and kaon masses which were computed on a small subset of configurations. 
The central values of the masses on the full set 
differ mildly from the ones estimated initially. As a result  the 
value for $q^2$ may differ significantly from zero. 
We corrected for this small mistuning as follows: from prior studies~\cite{Boyle:2007qe} 
we know that the momentum dependence of the form factor is well described by
a pole ansatz, $f_+^{K\pi}(q^2)=f_+^{K\pi}(0)/(1+q^2/M^2)$ with $M$ the pole mass. 
The latter can be determined from the data for the form factor 
computed with our choice of twist angle supplemented with the result
at $q^2_{\rm max}=(m_K-m_\pi)^2$, i.e. with the pion and kaon at rest.  We then
inter- or extrapolate to $q^2=0$ and use the corrected data in all subsequent 
steps of the analysis. Using instead a linear ansatz for the correction
leads to a numerically small difference 
far below the statistical error. This
dependence on the parameterisation can therefore safely be neglected.

Table~\ref{tab:ffresults} shows all accumulated simulation results and in figure~\ref{fig:final}
we provide a first visual impression of the data.
Both the table and the plot also show the data previously analysed 
in~\cite{Boyle:2007qe,Boyle:2010bh,Boyle:2013gsa}
where, at variance with what is done here, the geometric mean of the renormalisation 
constants $\sqrt{Z_V^\pi Z_V^K}$ was considered and data for the scalar current matrix element
was not considered at all. 
As can be seen from table~\ref{tab:ffresults} the renormalisation constants $Z_V^\pi$ and 
$Z_V^K$ differ significantly on all ensembles 
leading to differing results for the vector current matrix element. 
The form factors renormalised with either choice of $Z_V$ differ by mass dependent
cutoff effects and hence follow two independent trajectories in the approach to the
continuum limit where they are expected to agree by universality.

\section{Corrections towards the physical point}\label{sec:Corrections}
In order to compare our results to experimental measurements of the semi-leptonic 
kaon decay the data needs to be interpolated to physical quark masses,
extrapolated to the continuum limit and finite volume corrections  
estimated. There are further systematic effects which we will discuss later.

The precise physical process which we are considering is the decay of a 
neutral kaon into a charged pion, $K^0\to\pi^-$. While our sea quark masses
represent isospin symmetric QCD we approximate the situation found in nature
by carrying out a small interpolation of the data in the pion and kaon mass 
to their values
$m_{\pi^-}=139.6$MeV and $m_{K}=496.6$MeV~\cite{Agashe:2014kda}. 
 The results of this interpolation will then
be subject to a continuum extrapolation.

An  ansatz for the continuum-extrapolation, mass-interpolation and 
infinite-volume limit is of the generic form
\begin{equation}\label{eq:generic_fitansatz}
 f_+^{K\pi}(q^2=0,L,a^2,m_\pi^2,m_K^2)=A+f_{\rm NLO}(L,a^2,m_\pi^2,m_K^2)
			+f_{\rm NNLO}(L,a^2,m_\pi^2,m_K^2)+\dots\,.
\end{equation}
The subscripts NLO and NNLO indicate an expansion equal or similar in 
spirit to chiral perturbation theory augmented by additional terms that
parameterise the cutoff and volume-dependence of the lattice data.
The case $A=1$ corresponds
to the $SU(3)$-symmetric point of the form factor and there are no 
contributions $\propto(m_K^2-m_\pi^2)$ (Ademollo-Gatto~\cite{Ademollo:1964sr}). 

In a first attempt at finding a suitable model for eqn.~(\ref{eq:generic_fitansatz}) we try to parameterise
the simulation results at constant lattice spacing and defer the extrapolation to the continuum limit to later
in this section.
In particular, we consider four ans\"atze for the mass dependence,
\begin{equation}\label{eq:fitansaetze}
 \begin{array}{lrcll}
{\rm fit\,}\mathcal{A}:\,\,&f_+^{K\pi}(q^2=0,m_\pi^2,m_K^2)&=&1+f_2(f,m_\pi^2,m_K^2)\,,\\[2mm]
{\rm fit\,}\mathcal{B}:\,\,&f_+^{K\pi}(q^2=0,m_\pi^2,m_K^2)&=&1+f_2(f,m_\pi^2,m_K^2)+A_1{(m_K^2+m_\pi^2)(m_K^2-m_\pi^2)^2}\,,\\[2mm]
{\rm fit\,}\mathcal{E}:\,\,&f_+^{K\pi}(q^2=0,m_\pi^2,m_K^2)&=&A+\Delta M^2 A_0\,,\\[2mm]
{\rm fit\,}\mathcal{F}:\,\,&f_+^{K\pi}(q^2=0,m_\pi^2,m_K^2)&=&A+\Delta M^2 \left(A_0+A_1(m_K^2+m_\pi^2)\right)\,,
 \end{array}
\end{equation}
where $\Delta M^2\equiv (m_K^2-m_\pi^2)^2/m_K^2$. We have studied these ans\"atze 
previously in~\cite{Boyle:2013gsa}. They model
the basic properties of chiral perturbation theory at NLO~\cite{Ademollo:1964sr,Gasser:1984ux} 
and NNLO~\cite{Bijnens:2003uy}, respectively.
Fits
$\mathcal{A}$ and $\mathcal{B}$, which we will refer to as fits based on NLO
chiral perturbation theory, adopt the NLO-term~\cite{Gasser:1984ux},
\begin{equation}\label{eq:f2}
	\begin{array}{c}
	f_2(f,m_\pi^2,m_K^2,m_\eta^2)=\frac 32
		H(f,m_\pi^2,m_K^2)+\frac 32 H(f,m_\eta^2,m_K^2)\,,\\[2mm]
		{\rm with}\,\,	H(f,m_P^2,m_Q^2)=-\frac 1{64\pi^2f^2}
		\left(
		m_P^2+m_Q^2+2\frac{m_P^2 m_Q^2}{m_P^2-m_Q^2}
		\log\left(\frac {m_Q^2}{m_P^2}\right)\right)\,,
\end{array}
\end{equation}
where we will employ the leading order relation $m_\eta=\sqrt{(4m_K^2-m_\pi^2)/3}$ 
for the $\eta$-mass.
In fit $\mathcal{E}$ and $\mathcal{F}$ we replace $f_2$ by its Taylor expansion
in $m_K^2-m_\pi^2$. The latter class of fits will be referred to as
fits based on polynomial models. In fits $\mathcal{B}$ and $\mathcal{F}$ 
the term proportional to $A_1$  models the the mass dependence  
expected in the NNLO-term in chiral perturbation theory~\cite{Bijnens:2003uy}.

In the $SU(3)$-symmetric limit the form factor is constrained to unity at $q^2=0$, 
$f_+^{K\pi}(0)=1$. This relation 
is exactly reproduced at finite lattice spacing and in finite volume 
when computed from the matrix element of the vector current renormalised with
$Z_V^\pi$ or $Z_V^K$ as defined in eqn.~(\ref{eq:zv}).
The statistical error on the form factor therefore tends
towards zero when approaching this limit as can be seen from table~\ref{tab:ffresults}. 
Fits are therefore mostly constrained by data points further away from the experimental 
value of $\Delta M^2$ but close to the $SU(3)$-symmetric limit. While this can be a virtue 
if one knows the correct functional form of~(\ref{eq:generic_fitansatz})
it  constitutes a danger when one
only has a phenomenological model at hand. 
We note that the matrix element of the scalar current does not share this property at finite 
lattice spacing since cutoff effects can contribute when relating it
 to the vector current in the discretised theory.
In fact, the statistical signal of the form factor as obtained from the scalar current
deteriorates compared to the signal we observe for the vector current in the vicinity of the
$SU(3)$-symmetric point. We decided to include data from the scalar current only close
to the physical point 
where the statistical properties are similar to the ones from the vector current.

\subsection{Fits based on NLO chiral perturbation theory}
\begin{figure}
 \begin{center}
  \includegraphics[width=7.5cm]{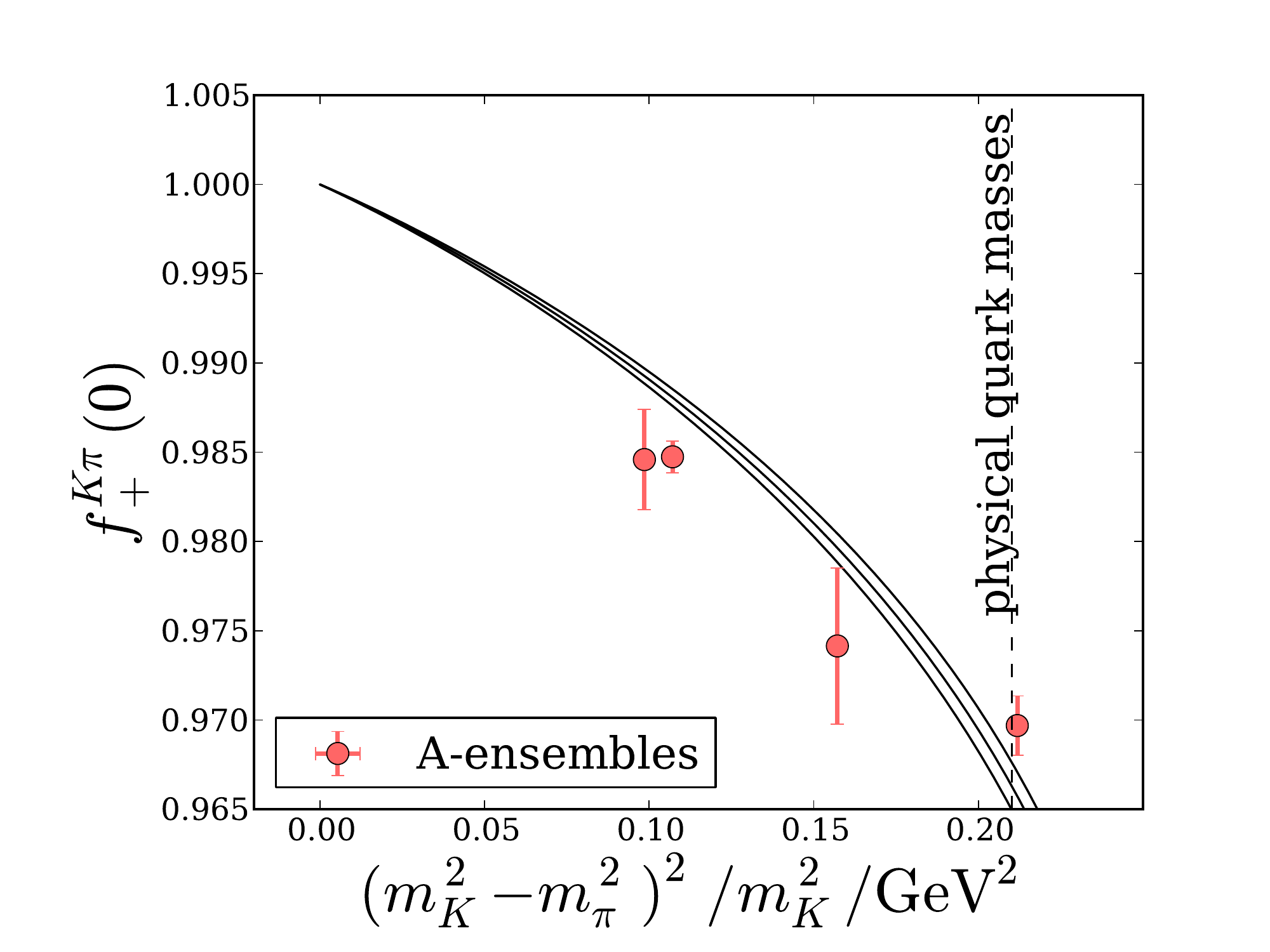}\hspace{-3mm}
  \includegraphics[width=7.5cm]{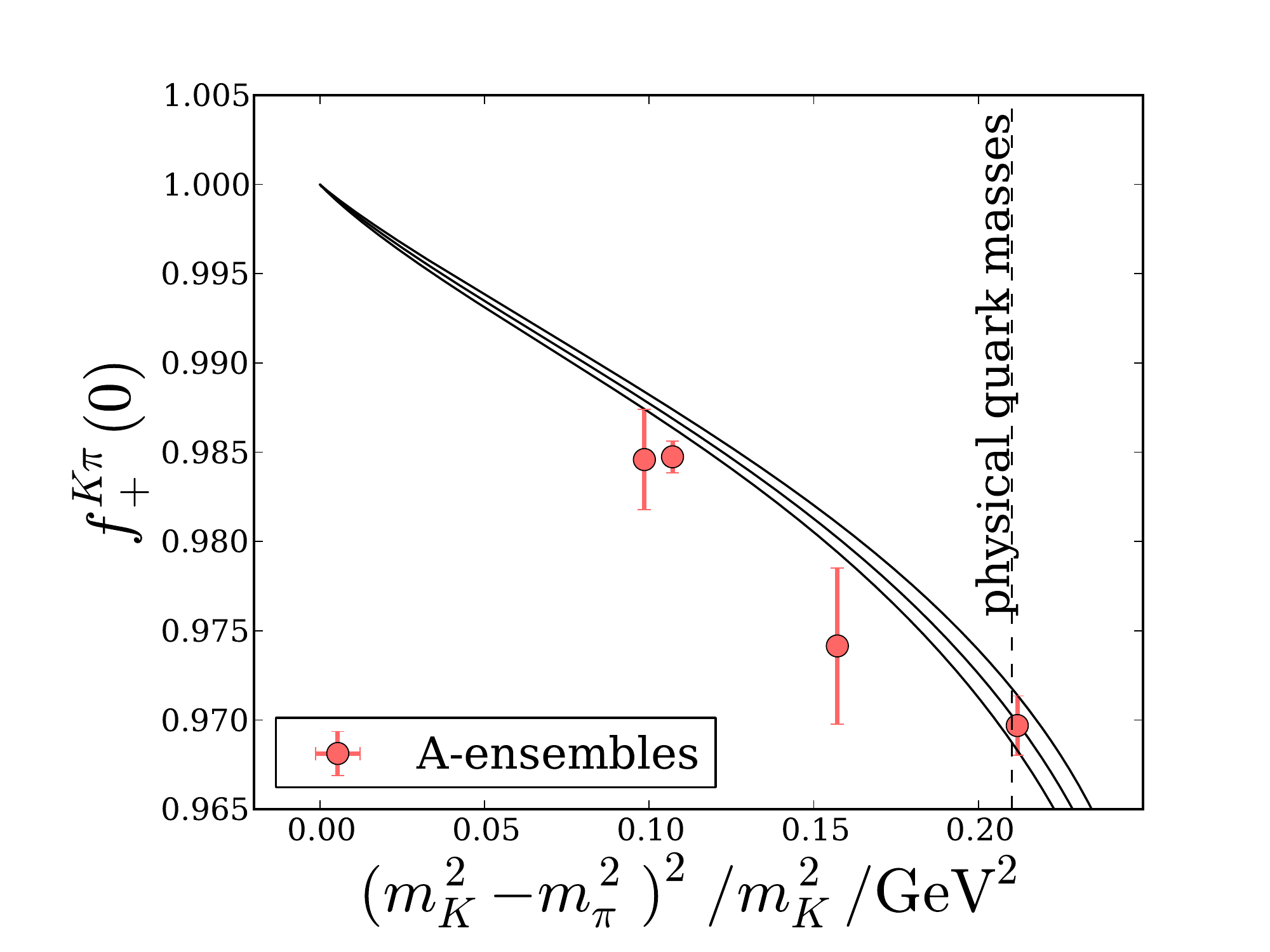}
 \end{center}
\caption{Illustration for fit $\mathcal{A}$ (left) and $\mathcal{B}$ (right)
to data for the form factor renormalised with $Z_V^\pi$
from the A-ensemble with a mass cut-off $m_\pi\approx 450$MeV. 
}\label{fig:ChPT_fits}
\end{figure}

Here we consider fits $\mathcal{A}$ and $\mathcal{B}$ where 
the decay constant $f$ (and in the latter case also $A_1$) are the parameters
to be determined. 
In figure~\ref{fig:ChPT_fits} we show representative fits to the form factor 
renormalised with $Z_V^\pi$ to ensembles A and C, respectively. 
By simple visual inspection and also following from the observation of 
large values of $\chi^2/\mathrm{dof}$
of 4.6 and 1.7, respectively, these ans\"azte are at variance with 
the simulation data. The situation is qualitatively the same when 
fitting ensembles C, when fitting the other two definitions  of the form factor
(vector current renormalised with $Z_V^K$ and scalar current)
and also when varying the mass cut-off in any of these cases.
Adding a model for the NNLO mass dependence in fit $\mathcal{B}$ 
improves the situation slightly. Large $\chi^2/\mathrm{dof}$ values however indicate
 a strong tension between the ansatz and the data.

Based on these findings we discard fits $\mathcal{A}$ and $\mathcal{B}$ 
from the further analysis.
While the above data strongly suggests that NLO chiral perturbation theory is 
unable to describe our results in the range between the $SU(3)$-symmetric point 
and the physical quark mass point,
it is still possible that the data is better described by the full
NNLO-expression~\cite{Bijnens:2003uy}. Our data set is however too limited  to allow
for a study of NNLO fits independent of experimental or model input.
\subsection{Fits based on polynomial models}
\begin{table}[htpb]
\small
\begin{center}
\begin{tabular}{l@{\hspace{1.5mm}}l@{\hspace{1.5mm}}l@{\hspace{1.5mm}}l@{\hspace{1.5mm}}l@{\hspace{1.5mm}}l@{\hspace{1.5mm}}l@{\hspace{1.5mm}}l@{\hspace{1.5mm}}l@{\hspace{1.5mm}}l}
\hline\hline\\[-4mm]
ME	&$\frac{m_\pi^{\rm cut}}{\rm MeV}$&$\fpzero^{\rm A}$&$\fpzero^{\rm C}$&$A^{\rm A}$&$A^{\rm C}$&$A_0^{\rm A}$\,GeV$^2$&$A_0^{\rm C}$\,GeV$^2$&$\frac{\chi^2}{\mathrm{dof}}$\\[1mm]
\hline
\multirow{4}{*}{$Z_V^\pi V$}&355&0.9703(16)&0.9689(16)&0.9970(52)&1.001(10)&-0.127(28)&-0.155(51)&0.37\\
			    &450&0.9704(16)&0.9687(16)&0.9994(23)&1.0002(26)&-0.138(17)&-0.150(17)&0.28\\
			    &600&0.9701(13)&0.9687(16)&0.99990(49)&1.0002(26)&-0.1416(80)&-0.150(17)&0.24\\
			    &700&0.97071(99)&0.9687(16)&0.999568(96)&1.0002(26)&-0.1373(49)&-0.150(17)&0.28\\
\hline                       
\multirow{4}{*}{$Z_V^K V$}  &355&0.9748(16)&0.9715(16)&0.9977(50)&1.0005(97)&-0.109(27)&-0.138(47)&0.28\\
			    &450&0.9748(16)&0.9714(16)&1.0027(24)&1.0017(25)&-0.133(17)&-0.144(16)&0.47\\
			    &600&0.9748(13)&0.9714(16)&1.00269(48)&1.0017(25)&-0.1327(78)&-0.144(16)&0.38\\
			    &700&0.97747(100)&0.9714(16)&1.001120(96)&1.0017(25)&-0.1125(50)&-0.144(16)&2.05\\
\hline                       
\multirow{2}{*}{$S$}	    &355&0.9715(14)&0.9717(19)&1.0022(80)&0.994(13)&-0.146(40)&-0.105(60)&0.00\\
			    &450&0.9715(14)&0.9716(19)&1.0022(80)&0.9890(68)&-0.146(40)&-0.083(35)&0.10\\
\hline \hline
\end{tabular}
\caption{Results for global fit $\mathcal{E}$ on ensembles A and C for various mass cut-offs. 
Both the coefficients $A$ and $A_0$
are allowed to depend on the cutoff.  The first column
indicates the form factor (vector matrix element (ME) renormalised with $Z_V^\pi$ or
$Z_V^K$, or scalar  ME).
 The superscript in the top line indicates the lattice
spacing for which this result was obtained (0.11fm for A and 0.08fm for C).}\label{tab:A-a-a}
\end{center}
\end{table}
\begin{table}
\small
\begin{tabular}{l@{\hspace{1.5mm}}l@{\hspace{1.5mm}}l@{\hspace{1.5mm}}l@{\hspace{1.5mm}}l@{\hspace{1.5mm}}l@{\hspace{1.5mm}}l@{\hspace{1.5mm}}l@{\hspace{1.5mm}}l@{\hspace{1.5mm}}l}
\hline
\hline\\[-4mm]
$\frac{m_\pi^{\rm cut}}{\rm MeV}$&$\fpzero^{\rm A}$&$\fpzero^{\rm C}$&$A^{\rm A}$&$A^{\rm C}$&$A_0^{\rm A}$\,GeV$^2$&$A_0^{\rm C}$\,GeV$^2$&$A_1^{\rm A}$\,GeV$^4$&$A_1^{\rm C}$\,GeV$^4$&$\frac{\chi^2}{\mathrm{dof}}$\\[1mm]
\hline
450&0.9697(16)&0.9710(38)&0.9990(77)&1.012(26)&-0.143(17)&-0.08(14)&0.02(12)&-0.41(92)&0.59\\
600&0.9697(16)&0.9710(38)&1.0000(15)&1.012(26)&-0.145(15)&-0.08(14)&0.001(56)&-0.41(92)&0.40\\
700&0.9695(15)&0.9710(38)&0.99946(16)&1.012(26)&-0.148(12)&-0.08(14)&0.020(23)&-0.41(92)&0.33\\
800&0.9695(15)&0.9710(38)&0.99946(16)&1.012(26)&-0.148(12)&-0.08(14)&0.020(23)&-0.41(92)&0.33\\
\hline
450&0.9740(16)&0.9740(37)&0.9986(75)&1.015(25)&-0.138(18)&-0.07(13)&0.08(12)&-0.49(88)&0.80\\
600&0.9739(16)&0.9740(37)&1.0021(15)&1.015(25)&-0.142(15)&-0.07(13)&0.027(57)&-0.49(88)&0.61\\
700&0.9734(15)&0.9740(37)&1.00067(16)&1.015(25)&-0.151(12)&-0.07(13)&0.079(23)&-0.49(88)&0.72\\
\hline
\hline
\end{tabular}
\caption{Results for fit $\mathcal{F}$ for ensembles A and C for various mass cut-offs. 
The superscripts on the coefficients $A$, $A_0$ and $A_1$ indicate the set of ensembles
for which this coefficient was determined. The first block of data  is for the form 
factor renormalised with $Z_V^\pi$ and the next block for the form factor
renormalised with $Z_V^K$. }
\label{tab:B-a-a-a}
\end{table}

\begin{figure}
 \begin{center}
  \includegraphics[width=7.5cm]{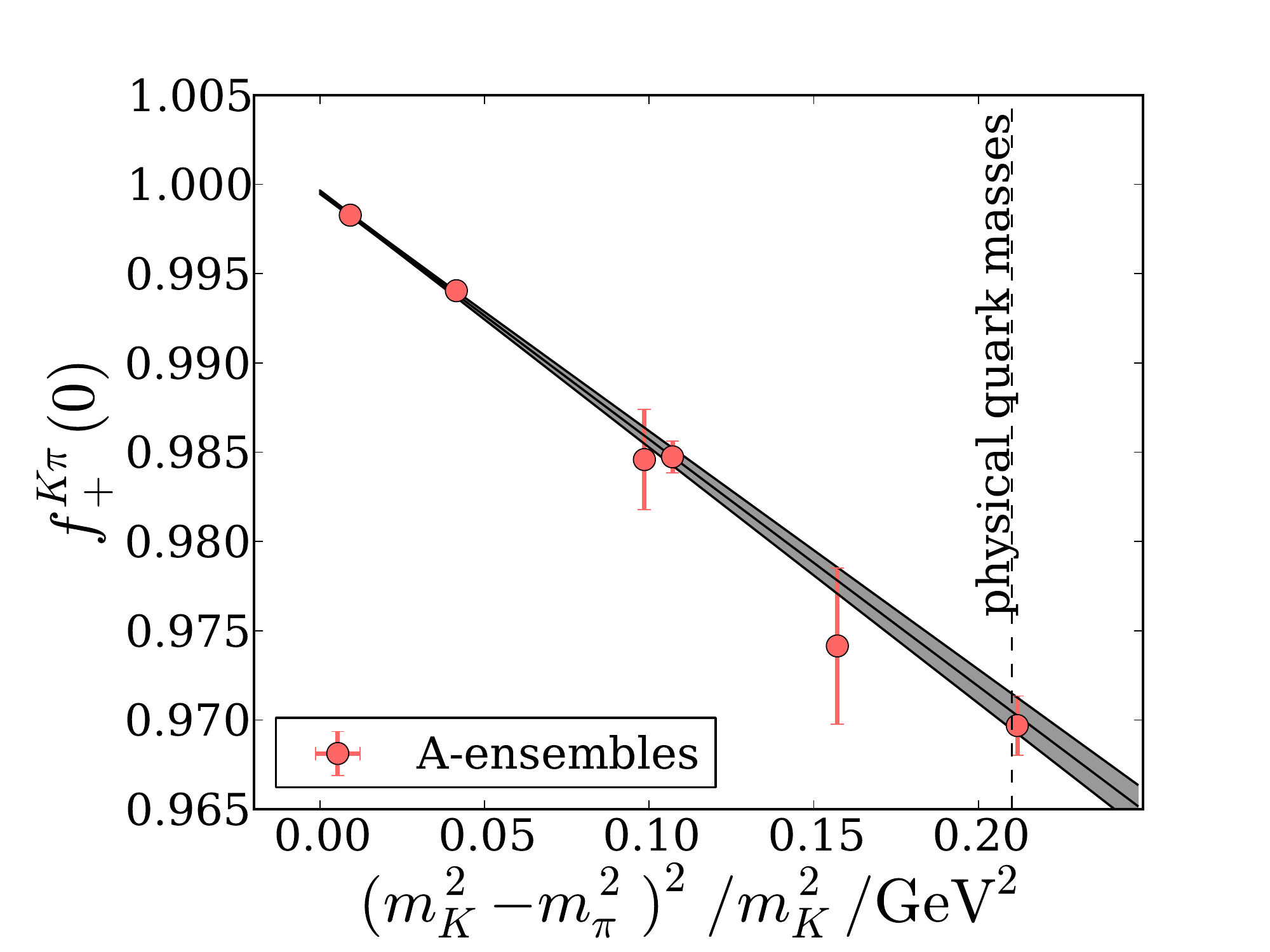}\hspace{-6mm}
  \includegraphics[width=7.5cm]{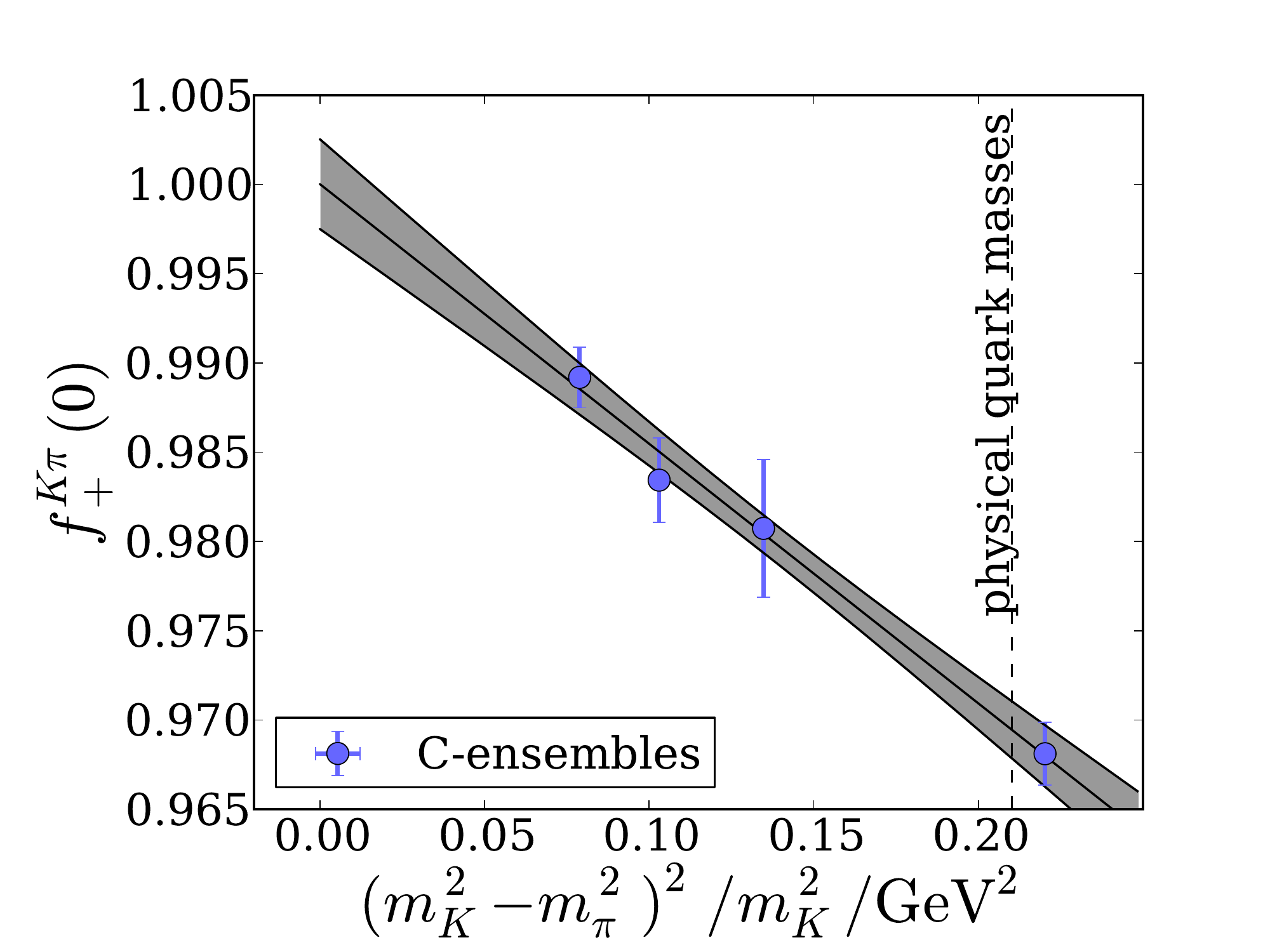}
 \end{center}
\caption{Illustration for fit $\mathcal{E}$
to data for the form factor renormalised with $Z_V^\pi$. The l.h.s. plot shows
a fit to all data of the A-ensembles; the r.h.s. plot shows the  fit to the C-ensembles.
}\label{fig:ChPT_fits3}
\end{figure}

\begin{figure}
 \begin{center}
  \includegraphics[width=7cm]{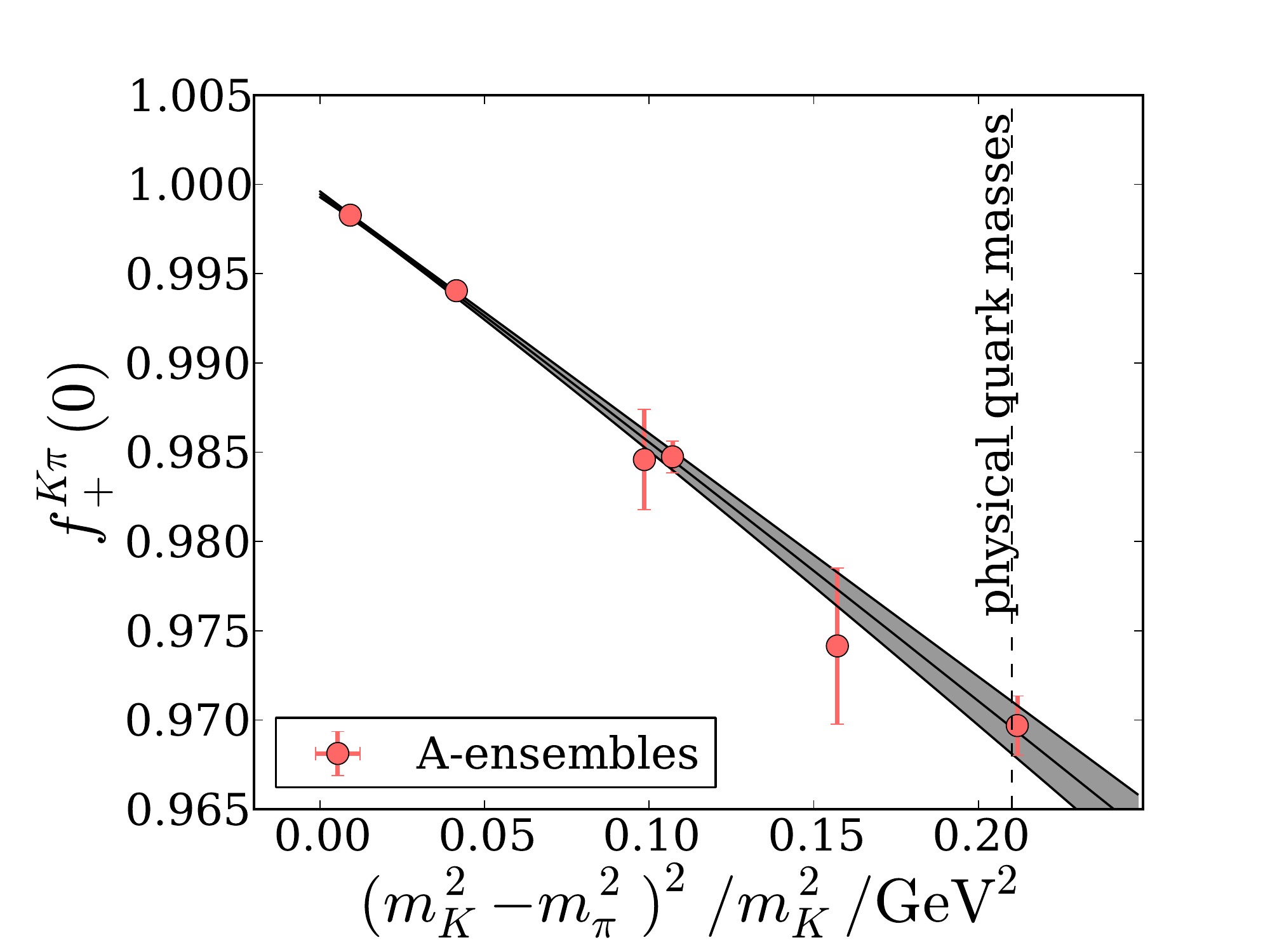}
  \includegraphics[width=7cm]{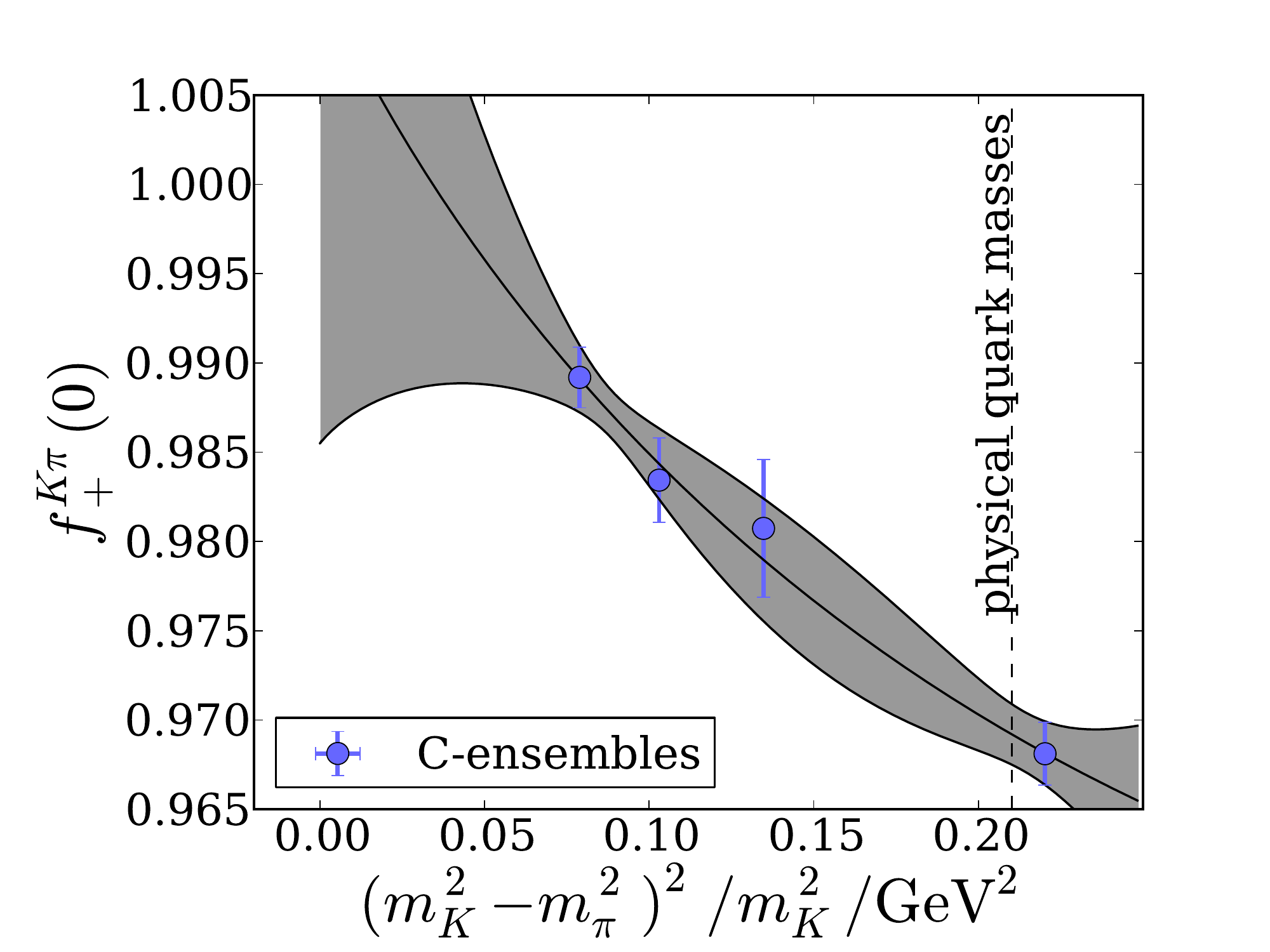}
 \end{center}
\caption{Illustration for fit $\mathcal{F}$
to all data for the form factor renormalised with $Z_V^\pi$. The l.h.s. plot
shows the fit to ensembles A, the r.h.s. plot the fit to ensembles C.
}\label{fig:ChPT_fits2}
\end{figure}

The results shown in figure~\ref{fig:final} show little  non-trivial structure in the
mass dependence of the form factor when plotted against $\Delta M^2$.
This suggests that a  polynomial model should work well in describing the
data.
We therefore repeat the study of the previous section with fits 
$\mathcal{E}$ and $\mathcal{F}$ which are simple polynomials in the
$SU(3)$-breaking parameter $\Delta M^2$. 
Some representative plots are shown in 
figures~\ref{fig:ChPT_fits3} and \ref{fig:ChPT_fits2} and numerical results for various
mass cut-offs are summarised in tables~\ref{tab:A-a-a}
and \ref{tab:B-a-a-a}. These fits turn out
to be of very good quality as indicated by the low value of $\chi^2/\mathrm{dof}$. 
In the fits with ansatz $\mathcal{E}$ the only sensitivity to 
corrections beyond the linear term (in $\Delta M^2$)
is indicated by the steep increase in $\chi^2/\mathrm{dof}$ when including the data 
closest to the $SU(3)$-symmetric point into the fit
with the form factor renormalised using $Z_V^K$. Indeed, in these cases
ansatz $\mathcal{F}$ performs better as can be seen in particular
by the reduced value of $\chi^2/\mathrm{dof}$ when fitting to the more comprehensive
set of ensembles A. We do not have enough data to allow for a meaningful
fit with ansatz $\mathcal{F}$ for the scalar form factor.

{
As illustrated in figure~\ref{fig:ChPT_fits3} (see also table~\ref{tab:A-a-a}),
the correction towards the physical point by means of fit~$\mathcal{E}$ constitutes
only a small shift with respect to the results on our physical pion
mass ensembles  A$_{\rm phys}$ and C$_{\rm phys}$. The statistical error is 
almost} {constant as long as the cut in the pion mass $m_\pi^{\rm cut}$
is chosen well below 600MeV.} 
{We conclude that
ensembles   A$_{\rm phys}$ and C$_{\rm phys}$ play a dominant role in 
determining the final result and the main role 
of the ensembles with heavier pion mass is   to 
determine the slope of the interpolation.}
%

\subsection{Continuum extrapolation}

{The results for the slope parameter $A_0$ in table~\ref{tab:A-a-a} 
are compatible between ensembles A and C. 
This indicates that we are not sensitive to a cut-off dependence
in the slope with $\Delta M^2$.
In fact, in the $O(a)$-improved theory we expect the lattice-spacing dependence of the
slope-parameter to be
\begin{equation}\label{eq:fitansaetze_a2}
 \begin{array}{lrcll}
  A_0(a)=A_0(0)\left(1+\alpha \left({a}{\Lambda_{\rm QCD}}\right)^2+...\right)\,,
 \end{array}
\end{equation}
with the parameter $\alpha$ parameterising the size of $a^2$ cut-off effects.
Assuming  conservatively
$\Lambda_{\rm QCD}=500$MeV, $(a\Lambda_{\rm QCD})^2$ is at most around 8\%. 
Based on these considerations the expected difference in $A_0$ between the
A and C ensembles is of $O(3\%)$ and hence smaller than the statistical error 
on $A_0$ itself.
Our observation that $A_0$ does not show any significant cutoff effects
is therefore not surprising. We confirmed that numerically the parameter $\alpha$ is compatible
with zero, in agreement with the observation that the results for
$A_0$ on ensembles A and C are compatible. 
}

\begin{figure}
 \begin{center}
  \includegraphics[width=10cm]{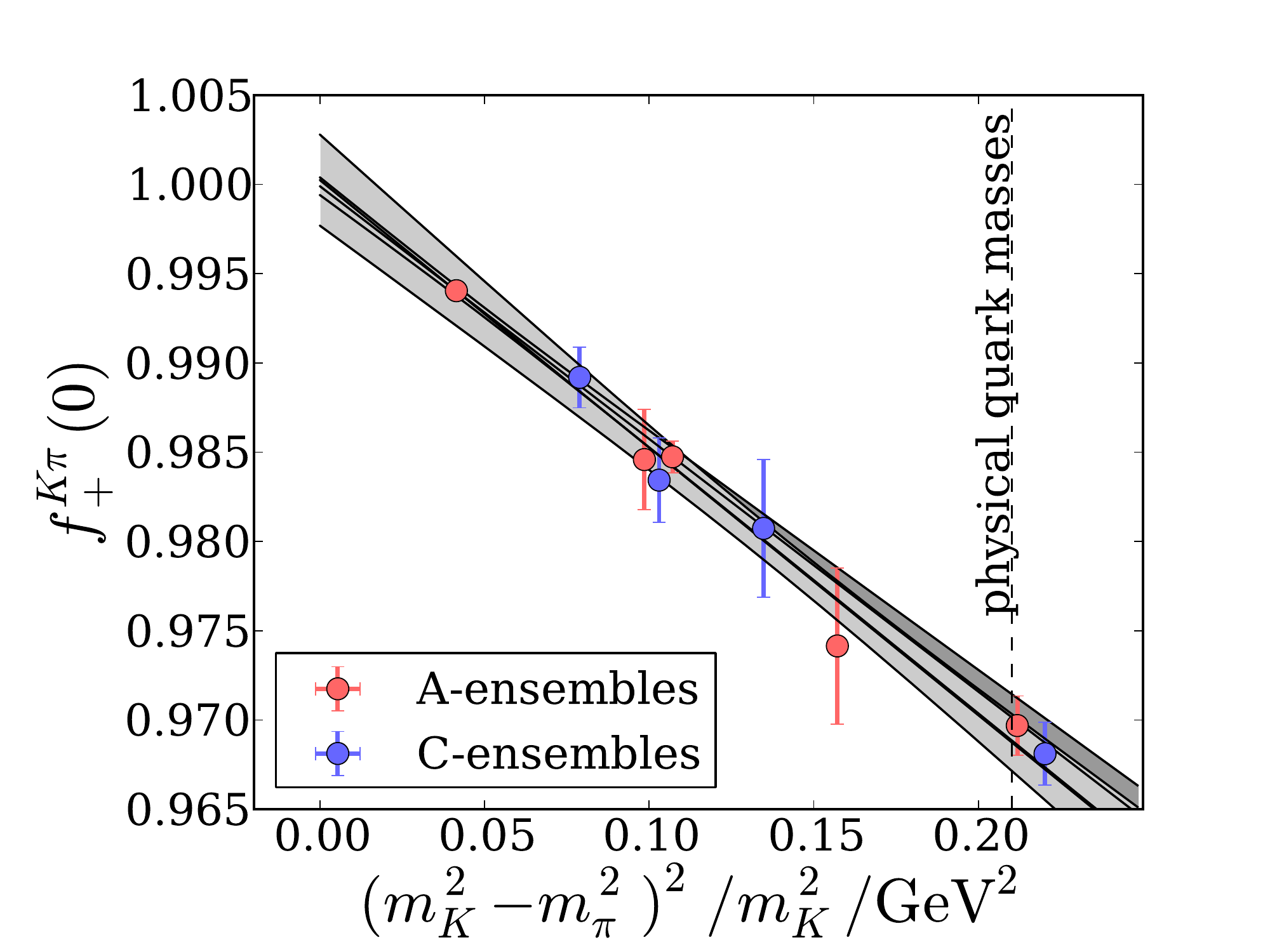}
 \end{center}
\caption{Illustration for fit $\mathcal{E}$
to all data for the form factor renormalised with $Z_V^\pi$. The coefficient $A_0$ 
is assumed to agree for ensembles A and C. Note the two sets of error bands, one
for ensemble A and one for ensemble C. 
}\label{fig:ChPT_fits_A-a-a}
\end{figure}

\begin{table}
\begin{center}
\begin{tabular}{l@{\hspace{1.5mm}}l@{\hspace{1.5mm}}l@{\hspace{1.5mm}}l@{\hspace{1.5mm}}l@{\hspace{1.5mm}}l@{\hspace{1.5mm}}l@{\hspace{1.5mm}}l@{\hspace{1.5mm}}l@{\hspace{1.5mm}}l}
\hline
\hline\\[-4mm]
$\frac{m_\pi^{\rm cut}}{\rm MeV}$&$\fpzero^{\rm A}$&$\fpzero^{\rm C}$&$A^{\rm A}$&$A^{\rm C}$&$A_0$\,GeV$^2$&$\frac{\chi^2}{\rm dof}$\\[1mm]
\hline
355&0.9701(15)&0.9690(16)&0.9982(45)&0.9970(52)&-0.134(24)&0.31\\
450&0.9699(12)&0.9691(14)&1.0002(17)&0.9994(19)&-0.144(12)&0.28\\
600&0.9699(12)&0.9692(12)&0.99999(46)&0.9993(15)&-0.1431(72)&0.23\\
700&0.97050(95)&0.9696(11)&0.999581(96)&0.9986(13)&-0.1383(48)&0.31\\
\hline\\[-4mm]
355&0.9745(15)&0.9717(15)&0.9991(43)&0.9962(50)&-0.117(23)&0.28\\
450&0.9743(12)&0.9718(14)&1.0035(17)&1.0009(20)&-0.139(12)&0.42\\
600&0.9745(12)&0.9721(12)&1.00280(44)&1.0004(15)&-0.1348(71)&0.38\\
700&0.97696(97)&0.9735(11)&1.001154(96)&0.9977(13)&-0.1151(48)&2.23\\
\hline\\[-4mm]
355&0.9716(14)&0.9716(19)&0.9997(68)&0.9997(72)&-0.133(34)&0.17\\
450&0.9719(13)&0.9710(18)&0.9949(53)&0.9940(53)&-0.109(26)&0.55\\
\hline\hline
\end{tabular}
\caption{Results for fit $\mathcal{E}$ for ensembles A and C for various mass cut-offs. 
The superscripts on the coefficient $A$ indicates the set of ensembles
for which this coefficient was determined. The first four data lines are for the form 
factor as renormalised with $Z_V^\pi$ and the next four lines for the form factor
renormalised with $Z_V^K$. The bottom two lines are for the results from the 
scalar matrix element.
}\label{tab:A-a-0}
\end{center}
\end{table}
Based on these findings we also attempt a modified version of fit~$\mathcal{E}$
which assumes the slope coefficient
$A_0$ to be independent of the lattice spacing. This fit is illustrated in
figure~\ref{fig:ChPT_fits_A-a-a} and the numerical results
are given in table~\ref{tab:A-a-0}. The plot shows two error bands for 
ensembles A and C, respectively. The fit is better constrained from ensembles A
owing to the larger range of quark masses for which simulation data is available. 
The fit is of very good quality as indicated by the  acceptable values of 
$\chi^2/\mathrm{dof}$ except for the largest mass cut-off in the case of the form factor
renormalised with $Z_V^K$.
{We note that when imposing cut-off independence of $A_0$ 
the statistical error on the result for the 
form factor reduces as $m_\pi^{\rm cut}$ is 
increased. We attribute this to the smaller statistical error on the
heavy pion mass ensembles which help constraining the fit. For the most
stringent pion mass cut $m_\pi^{\rm cut}=355$MeV, though, 
the statistical error is unaffected
by the assumption of cut-off independence of $A_0$. In this case 
the results on ${\rm A}_{\rm phys}$ and ${\rm C}_{\rm phys}$
dominate the fit-result.}
%

\begin{figure}
\begin{center}
 	\includegraphics[width=7cm]{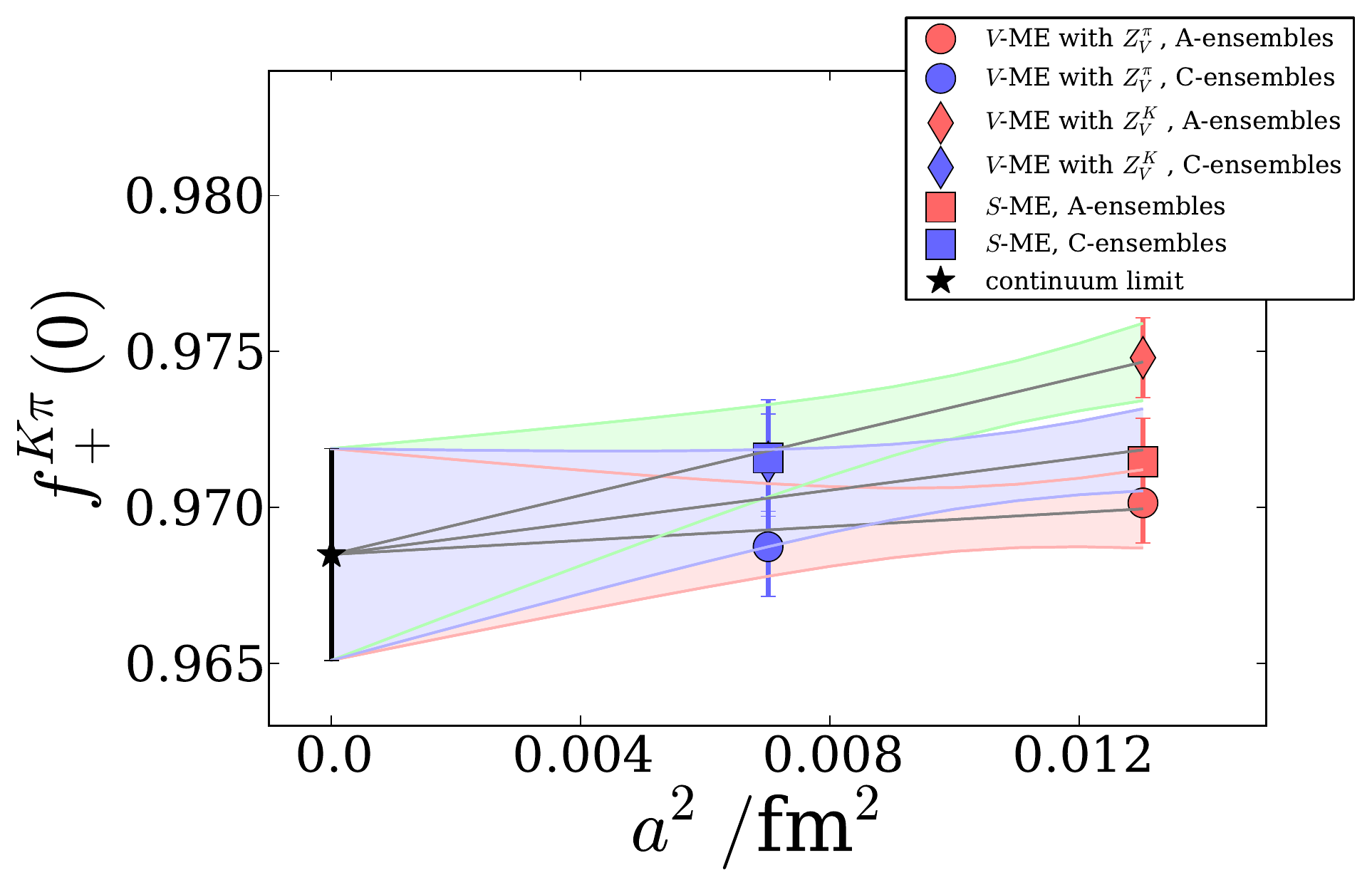}
 	\includegraphics[width=7cm]{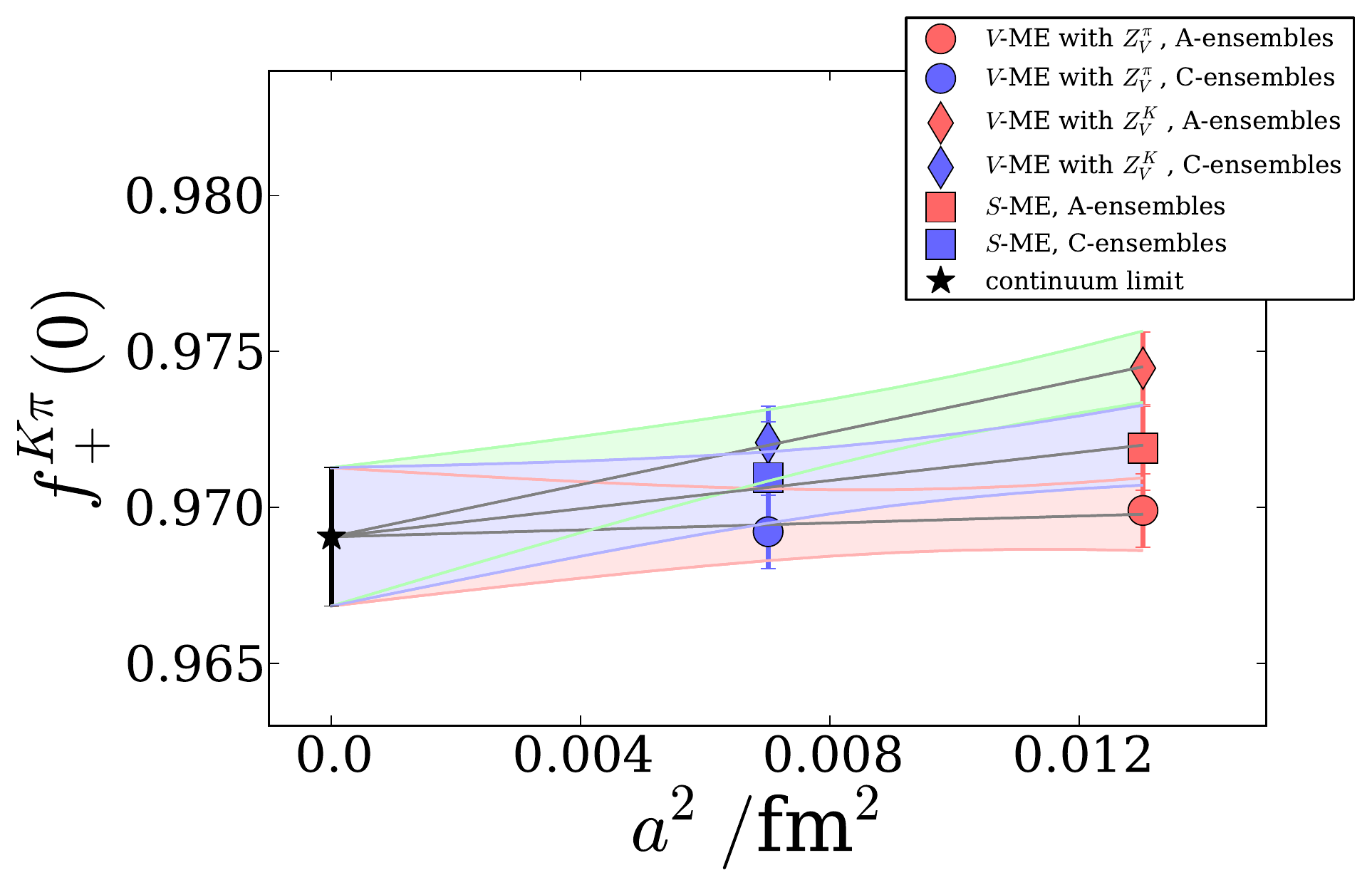}
\end{center}
\caption{Continuum extrapolation for results from fit $\mathcal{E}$ with mass cut-off 600MeV. Left: Coefficients $A$ and $A_0$ differ between ensembles A and C. Right: $A_0$ assumed to be the same for ensembles A and C.}\label{fig:massinterpolation}
\label{fig:cl}
\end{figure}

While our data would allow for taking three independent continuum limits for the form factors
as determined from the vector current renormalised with $Z_V^\pi$ and $Z_V^K$ and from the scalar current,
respectively, we instead analyse their 
joint continuum limit assuming universality: We impose that all three extrapolations have to
agree in the continuum limit. The combined extrapolation is shown in figure~\ref{fig:cl}
once without and once with the assumption of cutoff independence on $A_0$.
In table~\ref{tab:cl results} we only show fits for which the $\chi^2/\mathrm{dof}$ 
in the mass interpolation was below one.
The result is very stable under variation of the fit ansatz. To underline
the stability of our fit ansatz we also show the final result from fits $\mathcal{F}$ where
either $A_1$ or $A_0$ and $A_1$ are assumed to be cut-off independent.
The gain in statistical error from assuming $A_0$ to be cut-off independent carries over
to the continuum limit.
\begin{table}
\small
\begin{center}
\begin{tabular}{llllll}
\hline\hline\\[-4mm]
$m_\pi^{\rm cut}$/MeV	&355	&450	&600	\\
\hline &&&&&\\[-4mm]
global fit $\mathcal{E}$ 		    &0.9687(35)& 0.9685(34)& 0.9685(34)\\
global fit $\mathcal{E}$,  $A_0$ fixed	    &0.9690(33)& 0.9689(25)& 0.9691(22)\\
\hline&&&&&\\[-4mm]
global fit $\mathcal{F}$, $A_1$ fixed 	    &--        & 0.9683(35)& 0.9685(34)\\
global fit $\mathcal{F}$, $A_0, A_1$ fixed  &0.9694(34)& 0.9687(26)& 0.9690(22)\\
\hline\hline
\end{tabular}
\caption{Continuum limit results for the form factor based on fit $\mathcal{E}$ and
$\mathcal{F}$ for mass interpolation.}\label{tab:cl results}
\end{center}
\end{table}

\section{Systematic errors and final result}\label{sec:Systematics}
In this section we explain our choice for the final central value and present the 
full error budget. 

In the previous section we showed a number of approaches for the mass interpolation
and eventually combined it with the continuum extrapolation. Table~\ref{tab:cl results} 
summarises the results for the form factor in the continuum limit based on fits
with acceptable values of $\chi^2/\mathrm{dof}$. 
{First of all we see that all variants of the fit lead to 
compatible results. The major difference between the results is a reduction
in statistical error when $m_\pi^{\rm cut}$ is increased while assuming $A_0$
cut-off independent. As discussed in the previous section this
 originates from the comparatively small statistical error on the heavy pion ensembles.
Choosing $m_\pi^{\rm cut}=355$MeV the results from the three fits in table~\ref{tab:cl results}
are essentially the same. Without a clear preference we just pick fit $\mathcal{E}$ with
$A_0$ fixed as our final result.}

We now discuss the remaining sources of systematic errors:

\emph{finite volume errors:} 
{
The largest remaining source of systematic uncertainty is expected to be due to
finite volume effects (FVE). }
{
Since there are only single initial and final states in the $K\to\pi$ matrix element
we expect the dominant finite volume effects to be exponentially suppressed with 
$m_\pi L$. Moreover, $SU(3)$-symmetry enforces these effects to disappear 
exactly in the limit of equal quark masses, also in a finite volume. With the
smallest value of $m_\pi L=3.8$ on our ensembles we would therefore naively expect
finite volume effects on the form factor to be of order 
$\left(1-f_+^{K\pi}(0)\right)e^{-m_\pi L}=0.0007$. 
Based on~\cite{Ghorbani:2013yh} one obtains FVE about twice
this number on our ensembles ${\rm A}_{\rm phys}$ and 
${\rm C}_{\rm phys}$\footnote{Private communication.}. 
However, a conceptually clean calculation
of the finite volume effects in chiral perturbation theory including the 
effect of twisted boundary conditions (see~~\cite{Sachrajda:2004mi}) would be
 timely and useful in order to improve the quality of this estimate.
In the meantime we take twice our naive estimate as the systematic error due to
the finite volume.
}

\textit{partial quenching:}
	The strange quarks on ensemble ${\rm A}_{5}^{3}$ and on the C-ensembles 
        are partially quenched, i.e. the mass of the valence strange quark differs mildly from the 
	mass of the sea strange quark. 
	We checked that dropping $A_5^3$ from the analysis does not change the 
	outcome of our study in any significant way. On the C-ensembles the partial
	quenching is small and we do not expect significant systematic effects.

\textit{isospin symmetry:} The unitary light quarks in our simulations are isospin 
        symmetric. We approximate the isospin broken theory by 
        interpolating in the valence sector 
        to the value of $\Delta M^2$ corresponding to the physical point~\cite{Agashe:2014kda}.
 	This still leaves a systematic uncertainty due to the sea-quark isospin breaking
	which is difficult to quantify in our setup.
	We expect however that these effects are small compared to the other components of 
	our error budget. Techniques to include such effects in future calculations are being
  	developed~\cite{deDivitiis:2011eh,Tantalo:2013ty,Portelli:2013jla,Carrasco:2015xwa}.\\

These considerations lead to our final result:
\begin{equation}
 \fpzero = 0.9685(34)_{\rm stat}(14)_{\rm finite\, volume}\,.
\end{equation}
Using $|V_{us}|\fpzero=0.2163(5)$, as determined 
in~\cite{Antonelli:2010yf} from experiment in a phenomenological analysis,
we also predict
\begin{equation}
 |V_{us}|=0.2233(5)_{\rm experiment}(9)_{\rm lattice}\,,
\end{equation}
where the errors are from experiment and from the lattice computation, 
respectively.
With further input for $|V_{ud}|=0.97425(22)$ from super-allowed nuclear
$\beta$-decay the unitarity test for the first row of the CKM matrix yields
\begin{equation}
 1-|V_{ud}|^2-|V_{us}|^2=0.0010(4)_{V_{ud}}(2)_{V_{\,us}^{\rm exp}}(4)_{V_{\,us}^{\rm lat}}=0.0010(6)\,,
\end{equation}
where we have neglected the contribution from $|V_{ub}|\approx 10^{-3}$.
\section{Discussion and conclusions}\label{sec:Conclusions}
Simulations of lattice QCD are now feasible with physical light quark masses. This
step~change in simulation quality leads to the reduction if not removal of 
the often dominant systematic uncertainty due to the chiral extrapolation. In this
paper we have demonstrated how this can be achieved in practice in the case of the
$K\to\pi$ form factor at vanishing momentum transfer. This is a phenomenologically
important quantity allowing for unitarity tests of the CKM matrix and therefore for
stringent constraints of beyond SM physics. Lattice QCD is the 
only first principles computational tool that can predict this form factor.
An important strategic decision that has been made is 
in which way to make use of our previous results for unphysically heavy light
quark masses. We have chosen an intermediate path, i.e. we have used the 
information from the heavier ensembles to correct for a small mistuning
in the average up- and down-quark mass and the strange quark mass to the
physical point. 
{Our choice of fit ansatz and fit range gives the  
result at the physical point the heaviest weight and uses  
earlier simulation results with heavier pion masses merely for guiding small 
corrections towards the physical point.
In this way any model dependence in the fit ansatz is reduced to a minimum. We note 
that by restricting the set of ensembles entering the fit less (i.e. including 
ensembles with heavier pion mass) the 
statistical error on our final result could have been reduced by around 30\%. This
would however have come at the cost of an increased model dependence which we find
difficult to quantify.}
The remaining dominant systematic is due to finite volume effects for which
we provide an estimate based on effective theory arguments.

Whilst having results with physical quark masses and in the continuum limit
represents a huge leap forward there are a number
of improvements which we wish to address in the future. As mentioned above the largest
systematic uncertainty is now the one due to finite size effects. We would like to understand
them better, for example within the framework of
partially twisted chiral perturbation theory 
{or by generating results for the form factor
on different volumes with otherwise constant simulation parameters (quark masses and lattice
spacing)}. Moreover, in the analysis of experimental 
results in~\cite{Antonelli:2010yf} several approximations were made when averaging 
the individual decay channels. In particular, electromagnetic and isospin breaking
effects were treated within chiral perturbation theory. With the
advances made in this work it seems appropriate to rethink this approach and try to
treat these effects in a fully nonperturbative fashion in lattice field theory.
For an overview of progress in this direction we direct the reader 
to~\cite{deDivitiis:2011eh,Tantalo:2013ty,Portelli:2013jla,Carrasco:2015xwa}.\\

\noindent{\bf Acknowledgements:}
{
}
The research leading to these results has received funding from the Euro\-pe\-an
Research Council under the European Union's Seventh Framework Programme
(FP7/2007-2013) / ERC Grant agreement 279757 and also from the
Leverhulme Research grant  RPG-2014-118. 
The authors gratefully acknowledge computing time granted through the STFC funded DiRAC facility (grants ST/K005790/1, ST/K005804/1, ST/K000411/1, ST/H008845/1).
PAB acknowledges support from STFC Grant ST/J000329/1,  CJ is supported by 
DOE grant AC-02-98CH10886(BNL),
 NHC, RDM, DJM and HY are supported in part by U.S. DOE grant DE-SC0011941. 
Critical to this calculation were the Blue Gene/Q computers at the Argonne
Leadership Computing Facility (DOE contract DE-AC02-06CH11357) as well as
the RIKEN BNL Research Center and BNL Blue Gene/Q computers at the
Brookhaven National Laboratory.

\bibliographystyle{JHEP}
\bibliography{KL3}

\end{document}